\documentclass[reprint,aps,ams,amsmath,prx,twocolumn,longbibliography]{revtex4-1}
\usepackage{graphics}
\usepackage{graphicx}
\usepackage{epsfig}
\usepackage{amsmath}
\usepackage{amssymb}
\usepackage{amsfonts}
\usepackage{bm}
\usepackage{color}
\usepackage{bbm}
\usepackage{hyperref}
\usepackage[normalem]{ulem}
\usepackage{comment}
\usepackage{soul}
\usepackage[OT1]{fontenc}

\begin{document}

\title{Lorenz ratios of diagonal and Hall conductivities in a Dirac electron liquid}

\author{Alex Levchenko}
\affiliation{Department of Physics, University of Wisconsin-Madison, Madison, Wisconsin 53706, USA}

\date{May 14, 2025}

\begin{abstract}
The magnetotransport properties of a two-dimensional electron liquid in graphene are analyzed in the hydrodynamic limit under isothermal conditions. It is shown that the Wiedemann-Franz law does not hold in this regime and that the Lorenz ratios constructed from the diagonal and Hall conductivities generally differ from each other. The breakdown of the Wiedemann-Franz law is most pronounced near charge neutrality, where the Lorenz ratio can exceed its nominal universal Sommerfeld value by an order of magnitude. The corresponding Lorenz ratios exhibit a nonmonotonic dependence on the magnetic field, which is caused by emergent magnetic friction arising in liquids lacking Galilean invariance and possessing finite intrinsic conductivity.
\end{abstract}

\maketitle

\section{Introduction} 

The semiclassical theory of electron transport in metals and semiconductors \cite{AKL:1956,*AKL:1957a,*AKL:1957b} successfully explains many observed galvanomagnetic and thermoelectric effects \cite{Abrikosov,Ziman,Pippard}. Despite the quantum degeneracy of the electron system at low temperatures, $T\ll E_{\text{F}}$, where $E_{\text{F}}$ is the Fermi energy, the semiclassical approximation remains valid. This is because the electron mean free path, $l$, exceeds the quantum de Broglie wavelength, $l\gg\lambda\sim p^{-1}_{\text{F}}$, where $p_{\text{F}}$ is the Fermi momentum~\footnote{Throughout this paper, we adopt natural units where Planck's constant and Boltzmann's constant are set to unity, $\hbar=k_{\text{B}}=1$.}. Therefore, the concept of a trajectory in phase space remains meaningful, allowing classical equations of motion to be applied in conjunction with the Boltzmann equation to compute kinetic coefficients. As a result, Matthiessen's rule follows, stating that the total resistivity is the sum of contributions from different scattering channels, such as electron-impurity and electron-phonon scattering. This approach, however, neglects quantum mechanical interference and correlation effects on conductivity. Nonetheless, in most cases, these effects introduce only perturbative corrections to the conductivity computed within the semiclassical approximation.

In the presence of an external magnetic field $H$, the semiclassical approach remains valid as long as the de Broglie wavelength is small compared to the radius of the electron's cyclotron orbit, $\lambda\ll r_{\text{c}}=cp_F/eH$. 
While this condition is always satisfied in the weak-field limit, the semiclassical approximation can also describe transport phenomena in strong fields, 
$\omega_{\text{c}}\tau_{\text{tr}}\gg1$, where $\omega_{\text{c}}=eH/mc$ is the cyclotron frequency and $\tau_{\text{tr}}$ is the transport scattering time. This is because the conditions $\lambda\ll r_{\text{c}}$ and $\omega_{\text{c}}\tau_{\text{tr}}\gg1$ remain compatible as long as $E_{\text{F}}\tau_{\text{tr}}\gg1$. Naturally, neglecting the quantization of electron motion in a magnetic field omits the magneto-oscillations superposed on the monotonic rise in resistivity.

One key result of semiclassical transport theory that has been scrutinized over the years is the validity of the Wiedemann-Franz (WF) law, which relates the electrical conductivity tensor $\sigma_{ij}(H)$ and the thermal conductivity tensor $\kappa_{ij}(H)$ in a magnetic field \cite{AKL:1956,*AKL:1957a,*AKL:1957b}:
 \begin{equation}\label{eq:WF}
\kappa_{ij}(H)=\frac{\pi^2T}{3e^2}\sigma_{ij}(H).
\end{equation}
This relation follows under fairly general assumptions about the electron dispersion and the form of the collision integral. The ratio of these conductivity tensors defines the Lorenz coefficient, 
$L_{ij}=\kappa_{ij}/T\sigma_{ij}$, which takes the universal Sommerfeld value $L_{\text{WF}}=\pi^2/3e^2$. Equation \eqref{eq:WF} also yields several testable predictions.
For example, it implies that the Lorenz ratio is identical for diagonal (Drude) and transverse (Hall) conductivities and that this ratio remains independent of the applied field. It also predicts that it is temperature independent. 

For simple metals at the lowest-temperatures in the residual resistance regime, Lorenz ratios have been cataloged in Ref. \cite{Kumar:1993}. The WF law holds remarkably well across a wide range of conducting systems, regardless of variations in carrier concentration and sample purity. 

In the Fermi liquid regime, resistivity typically follows $\rho=\rho_0+AT^2$, where $\rho_0$ is the residual resistivity due to disorder, and the $AT^2$ term arises from electron-electron scattering. The coefficient $A$ can be linked to the Sommerfeld coefficient of the 
$T$-linear term in the electronic specific heat via the Kadowaki-Woods scaling. The electronic thermal resistivity times temperature, defined as $\varrho=L_{\text{WF}}T/\kappa$, is also known to exhibit a quadratic temperature dependence: $\varrho=\varrho_0+BT^2$. The validity of the WF law at zero temperature implies that $\rho_0$ and $\varrho_0$ must be equal. However, any finite temperature generally leads to a deviation from this law whenever $A\neq B$. In fact, in many systems, it is found that $B\gg A$, see e.g. discussion in Ref. \cite{Gourgout:2024}.
Additionally, numerous examples exist where the WF law is strongly violated, particularly in conductors with strong electron correlations \cite{Taillefer:2000,Tanatar:2007,Taillefer:2016,Jaoui:2018,Gooth:2018,Behnia:2020,Behnia:2023,Sunko:2024}. Theoretical explanations for these violations remain challenging, as no well-established framework exists for deriving thermoelectric transport coefficients in correlated metals. Nevertheless, several insightful phenomenological and microscopic calculations have been reported \cite{Li:2002,Mahajan:2013,Balents:2017,Li:2018,Levchenko:2020,Huang:2021,Tulipman:2023}.

Physically, the WF law is expected to hold if electron scattering is elastic. In this case, the relaxation times for charge and thermal currents coincide. It was recognized early on that this condition is met at very low temperatures, where elastic electron-impurity collisions dominate, and at temperatures well above the Debye temperature, where electron-phonon scattering becomes quasielastic, allowing the relaxation-time approximation to be applied. In contrast, at intermediate temperatures, electron scattering is predominantly inelastic, leading to significantly different relaxation times for charge and thermal transport, thereby causing the WF law to break down.

From this perspective, one should expect the WF law to be significantly modified in the hydrodynamic regime of an electron liquid \cite{Gurzhi:1968}, which emerges in high-mobility systems at intermediate temperatures. Indeed, the temperature must be sufficiently high to ensure a short electron-electron mean free path, yet not so high that momentum- and energy-relaxing collisions occur over much longer length scales. In this regime, electrons form a strongly correlated fluid, where transport is dominated by frequent collisions. 

This expectation was experimentally confirmed in graphene devices, where the observed Lorenz number exceeded its universal value in the Drude regime by an order of magnitude~\cite{Crossno:2016}. This striking result has attracted significant theoretical interest \cite{Principi:2015,Lucas:2016,Foster:2016,Lucas:2018,Zarenia:2019,Li:2020,Li:2022,Tu:2023a,Tu:2023b}. The underlying mechanism for such a strong deviation stems from the unique properties of electrons in graphene near charge neutrality, where charge and heat transport effectively decouple. While charge transport is constrained by the collision-limited intrinsic conductivity \cite{Kashuba:2008,Fritz:2008}, heat flow is convective and enhanced by the high entropy density, leading to a significantly increased effective thermal conductivity \cite{Lucas:2016,Li:2020}.    

Recent advances in measuring viscous magnetoresistance \cite{Zeng:2024} and magnetothermal transport \cite{Talanov:2024} in graphene devices highlight the need for additional theoretical insight, particularly regarding observed anomalies in the Lorenz ratio. In this work, we employ the magnetohydrodynamic framework for transport in weakly inhomogeneous graphene, developed in Refs. \cite{Narozhny:2015,Schmalian:2017,Levchenko:2024}, to derive the density and field dependence of the Lorenz ratio for both the longitudinal and Hall components of the conductivity tensors.

The status of the WF law in graphene subjected to a magnetic field was previously analyzed theoretically in Refs. \cite{Lucas:2018,Tu:2023b} from two completely different perspectives. In Ref. \cite{Tu:2023b}, a Boltzmann-transport model with bipolar diffusion and an assumed energy gap induced by the substrate was considered. This study also employed an ad hoc model for the electron relaxation time, incorporating three different parameters corresponding to the scattering strength of short-range disorder, acoustic phonons, and long-range Coulomb disorder. This model predicts an increase in the peak height of the Lorenz ratio for the diagonal conductivities as a function of the magnetic field, as well as a sign change in the Lorenz ratio for the Hall conductivities depending on temperature. In contrast, the results of Ref. \cite{Lucas:2018} are based on a kinetic equation approach that incorporates the interplay between momentum-conserving electron-electron collisions and momentum-relaxing scattering on impurities, enabling the description of the ballistic-to-hydrodynamic crossover. It was shown that the strong violation of the WF law in this framework is primarily driven by electron interaction effects and that the Lorenz ratio in a finite magnetic field can be either smaller or larger than the WF law predicts, depending on the relationship between the cyclotron frequency and the scattering rates.  

In line with Ref. \cite{Lucas:2018}, we work within the regime of electron hydrodynamics in this paper. However, our approach goes beyond the perturbative treatment of collision operators, as we formulate hydrodynamic theory in a more general framework. This allows us to clearly elucidate crucial physics that was overlooked in previous studies on the WF law and the Lorenz ratio of a magnetized Dirac plasma. The electron fluid in graphene does not possess Galilean invariance; therefore, its dissipative properties are governed by viscosity and intrinsic conductivity. The latter plays a crucial role in magnetoresistance, as demonstrated in Ref. \cite{Levchenko:2024} and observed in Ref. \cite{Ponomarenko:2023}. Due to intrinsic conductivity, the magnetic friction arising from the slippage of the fluid relative to the field lines modifies the force-balance condition, ultimately altering all components of the thermoelectric resistance matrix. Consequently, the WF law and the Lorenz ratio are modified in a highly nontrivial manner.

\section{Hydrodynamic magnetotransport in graphene}

The hydrodynamic description of electron transport applies on length and time scales that are long compared to the mean free paths set by electron-electron collisions $l_{\text{ee}}$ but short compared to those associated with the relaxation of electron momentum and energy due to scattering off disorder $l_{\text{ei}}$ and phonons $l_{\text{ep}}$. These stringent conditions are typically satisfied only in high-mobility semiconductor heterostructures and graphene devices at intermediate temperatures. In this regime, hydrodynamic equations describe the evolution of the densities of particles $n$, energy $\epsilon$, and momentum $\bm{p}$, which are conserved in electron-electron collisions. In practice, it is often convenient to replace the evolution equation for energy density with an equivalent equation for the entropy density $s$ \cite{LL-V6}. 

In this section, we briefly recapitulate the main ingredients of the hydrodynamic theory applicable to electron systems in weakly inhomogeneous graphene devices, as developed in Refs. \cite{Li:2020,Levchenko:2024}. We then apply this theory to the calculation of the thermoelectric resistivity matrix in a magnetic field, with the goal of extracting the Lorenz ratios, which are the focal point of this study.

Consider a two-dimensional electron liquid in graphene subjected to a perpendicular magnetic field $\bm{H}$ and exposed to a random disorder potential with a correlation radius $\xi$ much larger than the electron mean free path, $\xi\gg l_{\text{ee}}$, yet small compared to the overall system size. In this setup, disorder averaging can be performed to obtain an effective hydrodynamic description that incorporates disorder-induced renormalizations and correlations with electron scattering. To have a concise description, we introduce the following column vector notations
\begin{equation}\label{eq:vectors}
\vec{x}=\left(\begin{array}{c}n \\ s \end{array}\right), \quad \vec{\bm{J}}=\left(\begin{array}{c}\bm{j}_n \\ \bm{j}_s \end{array}\right), \quad \vec{\bm{X}}=\left(\begin{array}{c} -e\bm{E} \\ \bm{\nabla}T \end{array}\right), 
\end{equation}  
for the particle and entropy densities $\vec{x}$, their respective currents $\vec{\bm{J}}$, and thermodynamically conjugated forces $\vec{\bm{X}}$. The column vector of forces consists of the electromotive force (EMF) and the temperature gradient. For a steady-state flow, the continuity equation for currents reduces to 
\begin{equation}\label{eq:currents}
\bm{\nabla}\cdot\vec{\bm{J}}=0,\quad \vec{\bm{J}}=\vec{x}\bm{u}-\hat{\Upsilon}\vec{\bm{X}}+\frac{e}{c}\vec{\gamma}[\bm{u}\times\bm{H}].
\end{equation} 
The first term in the current corresponds to the convective flow of the fluid with hydrodynamic velocity $\bm{u}$. The second term corresponds to currents relative to the fluid at rest, which is governed by the matrix of intrinsic thermoelectric coefficients $\hat{\Upsilon}$. 
The last term describes the contribution to the current from EMF caused by the Lorentz force exerted by a magnetic field in a moving liquid \cite{LL-V8}. 
The corresponding $2\times2$ dissipative matrix and column-vector of kinetic coefficients are defined as follows  
\begin{equation}
\hat{\Upsilon}=\left(\begin{array}{cc} \sigma/e^2 & \gamma/T \\ \gamma/T & \kappa/T \end{array}\right), \quad \vec{\gamma}=\left(\begin{array}{c} \sigma/e^2 \\ \gamma/T \end{array}\right). 
\end{equation}
Here $\sigma$ is the intrinsic conductivity, $\kappa$ is the intrinsic thermal conductivity, and $\gamma$ is the intrinsic thermoelectric coefficient. 

As shown in Ref. \cite{Li:2020}, an important effect of disorder is to renormalize the intrinsic conductivity, shifting it from its value in the pristine system:
\begin{equation}\label{eq:sigma}
\sigma=\sigma_0+e^2\chi, \quad \chi=\frac{1}{2\eta}\int\frac{d^2q}{(2\pi)^2}\frac{\langle|\delta n_{\bm{q}}|\rangle}{q^2}.
\end{equation}   
Here $\eta$ denotes the shear viscosity of the Dirac fluid. For monolayer graphene, a weak-coupling analysis yields the intrinsic conductivity as \cite{Kashuba:2008,Fritz:2008}
\begin{equation}
\sigma_0=\frac{e^2}{2\pi\alpha^2_T}, \quad \alpha_T=\frac{\alpha_g}{1+\frac{\alpha_g}{4}\ln(\Lambda/T)}, 
\end{equation}
where $\alpha_g=e^2/\varepsilon v$ is the bare dimensionless interaction constant, with $\varepsilon$ being the dielectric constant and $v$ the band velocity of graphene. A high-energy cutoff $\Lambda$ is introduced to regularize the renormalization group procedure.
The second term in Eq. \eqref{eq:sigma} accounts for an increase in the effective conductivity beyond its intrinsic value. This enhancement arises from disorder-induced fluctuations in the particle density, $\delta n$, around charge neutrality. For a model of encapsulation-induced long-range disorder, the spectral density of disorder potential fluctuations $U_{\bm{q}}$ remains weakly dependent on $q$ in the long-wavelength limit $q\to0$. In this case, density fluctuations can be related to disorder fluctuations via linear screening theory, $\delta n_{\bm{q}}/q=-U_{\bm{q}}/2\pi e^2$, 
so that the enhancement factor in Eq. \eqref{eq:sigma} takes the form
\begin{equation}
\chi=\frac{1}{2\eta}\frac{\langle U^2\rangle}{(2\pi e^2)^2}.
\end{equation} 
Near charge neutrality, dimensional analysis suggests that viscosity scales with the density of the thermal excitation cloud, 
$\eta\sim (T/v)^2$, a result supported by rigorous calculations \cite{Muller:2009}. Consequently, one generally expects $\sigma$ to exhibit significant temperature dependence. In the following, we will not impose any model-specific assumptions about the $T$-dependence of 
$\sigma$ and will instead treat it phenomenologically as an input parameter of the theory.

Disorder also renormalizes all other kinetic coefficients. However, due to particle-hole symmetry, the intrinsic thermoelectric coefficient vanishes at charge neutrality, $\gamma\to0$ as $n\to0$, making its renormalization at small density negligible.

A final remark on the matrix $\hat{\Upsilon}$ concerns its dependence on the magnetic field. Throughout this work, we assume that the cyclotron radius $r_{\text{c}}$ is larger than the electron-electron scattering length, which in graphene is parametrically of the same order as the thermal de Broglie wavelength, $l_{\text{ee}}\sim l_T\sim v/T$. In this regime, the field dependence of the intrinsic conductivities, as well as their respective Hall components, can be neglected. Thus, within this approximation, the elements of $\hat{\Upsilon}$ behave as scalar quantities.

The continuity equation for the momentum flux density can be presented in the form of the force balance condition \cite{Levchenko:2024}
\begin{align}\label{eq:force-balance}
&\vec{x}^{\mathbb{T}}\vec{\bm{X}}+\frac{1}{l^2_H}[\hat{\bm{z}}\times\vec{\gamma}^{\mathbb{T}}\vec{\bm{X}}]\nonumber \\ 
&=-(k+k_H)\bm{u}+\frac{n}{l^2_H}[\hat{\bm{z}}\times\bm{u}]+\eta\bm{\nabla}^2\bm{u},
\end{align} 
where $\hat{\bm{z}}$ is the unit vector along the field perpendicular to the plane of the 2D electron system and superscript $\mathbb{T}$ denotes vector transposition. We have also introduced the magnetic length $l_H=\sqrt{c/eH}$. 

The disorder averaging leading to Eq. \eqref{eq:force-balance} modifies the force-balance equation in several crucial ways. It leads to the friction force $-k\bm{u}$ with the corresponding coefficient whose magnitude is proportional to the disorder strength  
\begin{equation}
k=\frac{\langle(s\delta n-n\delta s)^2\rangle}{2\left(\frac{\sigma_0}{e^2}s^2-2\frac{\gamma_0}{T}ns+\frac{\kappa_0}{T}n^2\right)}.
\end{equation}
Noteworthy, unlike the friction force due to quench short-ranged disorder, this friction coefficient has rather complicated temperature and density dependence. Near charge neutrality, $n\to0$, it simplifies to $k=\frac{e^2}{2\sigma_0}\langle\delta n^2\rangle$. 
This frictional force competes with the magnetic friction which is described by the coefficient 
\begin{equation}
k_H=\frac{\sigma_0}{e^2l^4_H}.
\end{equation}
It should be noted that it only arises in systems with broken Galilean invariance as it explicitly depends on the intrinsic conductivity. The competing nature of these frictional effects should be understood as follows.  The disorder friction tries to equilibrate the fluid in the lab-frame thus bringing flow velocity to zero. In contrast, the magnetic friction tries to bring the liquid to the frame moving with the magnetic drift velocity $\bm{v}=c[\bm{E}\times\bm{H}]/H^2$. This can be readily seen by combining $-k_H\bm{u}$ term and the second term on the left-hand-side of Eq. \eqref{eq:force-balance} together resulting in $-k_H(\bm{u}-\bm{v})$. 

The remaining terms in Eq. \eqref{eq:force-balance} are standard. The last term on the right-hand-side is the viscous force, while the second term is the Lorentz force. The first term on the left-hand-side is the thermodynamic force due to pressure gradients which follows from the basic thermodynamic relation $\bm{\nabla}P=n\bm{\nabla}\mu+s\bm{\nabla}T$, where the gradient of the chemical potential should be absorbed into the EMF. 

Finally, Eq. \eqref{eq:force-balance} may include additional forces stemming from the Hall (odd) viscosity \cite{Avron:1998}. We neglect these terms, which adhears to the same accuracy as the neglect of field dependence of $\hat{\Upsilon}(H)$ in the semiclassical limit.

\begin{figure*}[t!]
\includegraphics[width=0.325\linewidth]{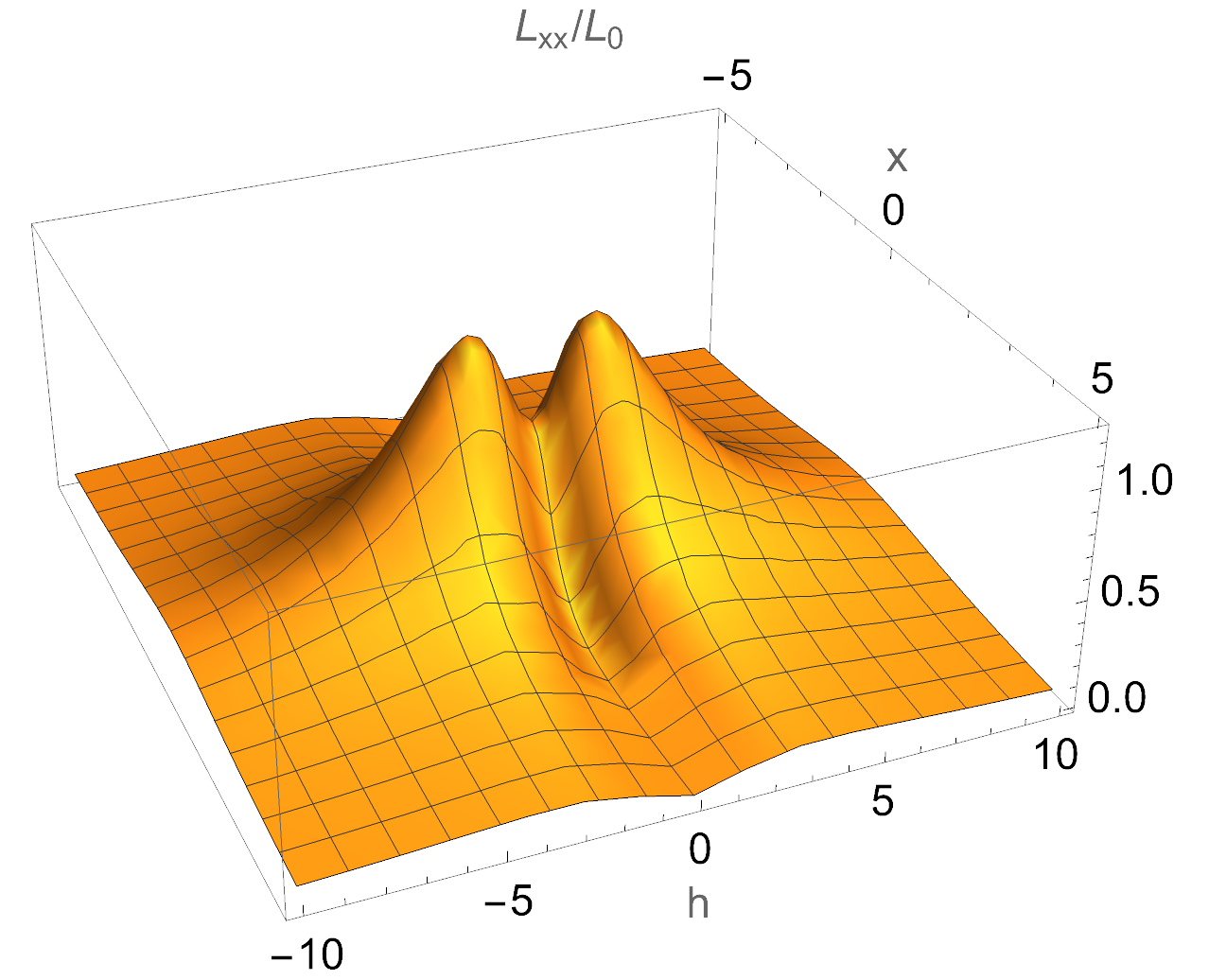}
\includegraphics[width=0.325\linewidth]{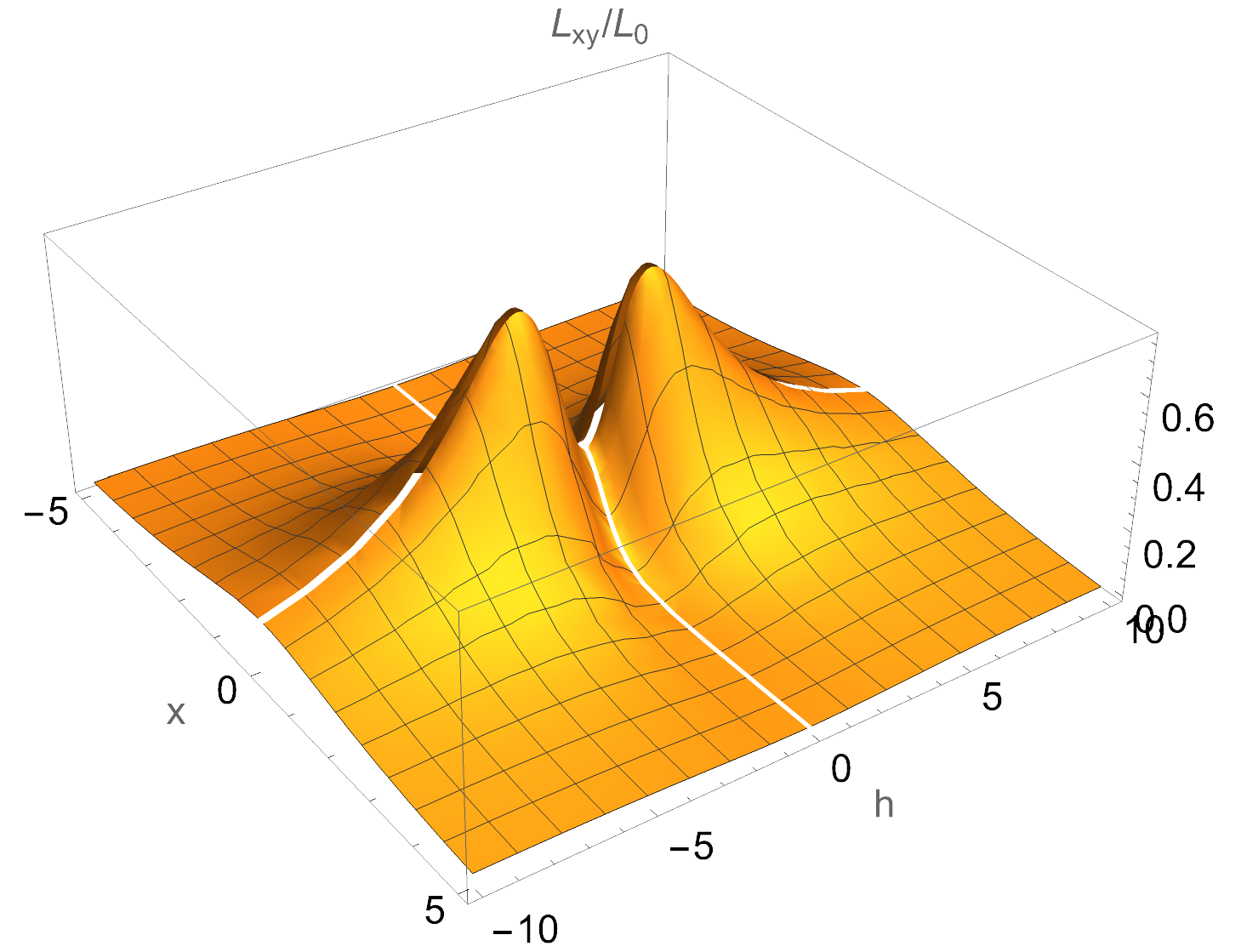}
\includegraphics[width=0.325\linewidth]{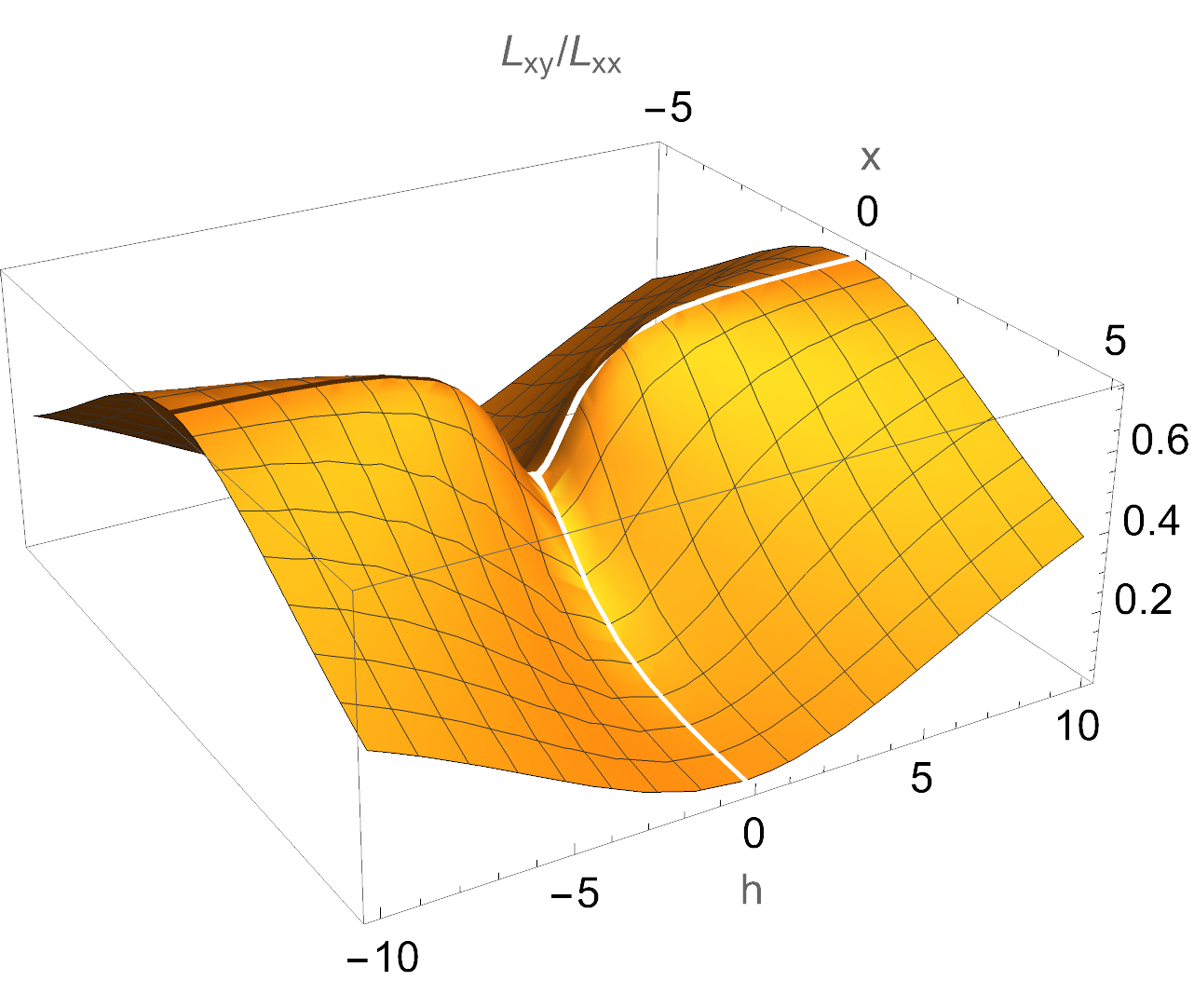}
\caption{Plots for the Lorenz ratios of the diagonal conductivities $L_{xx}(n,H)$, Hall conductivities $L_{xy}(n,H)$, and their relative ratio $L_{xy}/L_{xx}$ as a function of particle density $n$ and magnetic field $H$. Each plot is presented in the dimensional units $x\propto n$ and $h\propto H$ per convention specified in Eq. \eqref{eq:x-h} and normalization of $L_{ij}$ is chosen in the units of $L_0=s^2/\sigma_0k$ and $\chi_\sigma=0.25$.}\label{fig:L-3D}
\end{figure*}

For a given sample geometry, Eq. \eqref{eq:force-balance} should be solved with the appropriate boundary conditions. These typically include no-slip or no-stress conditions on the flow velocity, or a mixed condition with a finite slip length. This results in rather complex patterns of current distribution, as the response is nonlocal due to viscous terms. In the case of a Hall-bar geometry, some analytical solutions can be found e.g. Refs. \cite{Torre:2015,Levitov:2016,Alekseev:2018,Danz:2020}. In general, knowledge of the flow profile within the device allows for the calculation of currents using Eq. \eqref{eq:currents}, thereby establishing a linear relationship between the currents and the driving forces. This leads to an effective resistance matrix. The resulting thermoelectric resistances also depend on whether one considers isothermal or adiabatic transport, with the difference arising from the conditions imposed on nullifying gradients or fluxes (see \S 6.2 in Ref. \cite{Abrikosov} for a detailed discussion on this distinction).

To circumvent some of these complications, it is useful to observe that comparing the viscous term and the friction term in Eq. \eqref{eq:force-balance} introduces a characteristic length scale in the problem:
\begin{equation}
l_{\text{G}}=\sqrt{\eta/k},         
\end{equation}
which is known as the Gurzhi length \cite{Gurzhi:1968}. If the sample size $L$ exceeds $l_{\text{G}}$, deviations from uniform flow occur only near the sample boundaries. In this regime, viscous forces can be neglected in Eq. \eqref{eq:force-balance}, as relaxation is primarily governed by bulk friction. In this particular limit, Eq. \eqref{eq:force-balance} can be easily solved, yielding the result
\begin{equation}\label{eq:u}
\bm{u}=-\frac{(k+k_H)\bm{F}+(n/l^2_H)[\hat{\bm{z}}\times\bm{F}]}{(k+k_H)^2+(n/l^2_H)^2},
\end{equation}
where 
\begin{equation}
\bm{F}=\vec{x}\vec{\bm{X}}+\frac{1}{l^2_H}[\hat{\bm{z}}\times\vec{\gamma}^{\mathbb{T}}\vec{\bm{X}}]. 
\end{equation}
Using this result along with the expression for currents in Eq. \eqref{eq:currents} establishes a linear relationship between $\vec{\bm{J}}$ and $\vec{\bm{X}}$. The proportionality matrices define the corresponding effective thermoelectric magnetoresistances. The vector structure of the solution in Eq. \eqref{eq:u} naturally suggests the following matrix decomposition:
\begin{equation}\label{eq:J}
\vec{\bm{J}}=-\langle\hat{\Upsilon}\rangle\vec{\bm{X}}-\langle\hat{\Xi}\rangle [\hat{\bm{z}}\times\vec{\bm{X}}]. 
\end{equation}  
In these notations the brackets on top of matrices $\langle\hat{\Upsilon}\rangle$ and $\langle\hat{\Xi}\rangle$ denote the disorder-averaged quantities. For example, the diagonal magnetoconductivity is given by $\sigma_{xx}(H)=e^2\langle\hat{\Upsilon}\rangle_{11}$ whereas the Hall conductivity  $\sigma_{xy}(H)=e^2\langle\hat{\Xi}\rangle_{11}$. Thermal conductivities involve all matrix elements of $\langle\hat{\Upsilon}\rangle$ and $\langle\hat{\Xi}\rangle$ since they are defined under the condition of vanishing electrical current. This condition enforces a relation between the local values of the electric field and temperature gradients that mixes $\langle\hat{\Upsilon}\rangle$ and $\langle\hat{\Xi}\rangle$.  

\begin{figure*}[t!]
\includegraphics[width=0.325\linewidth]{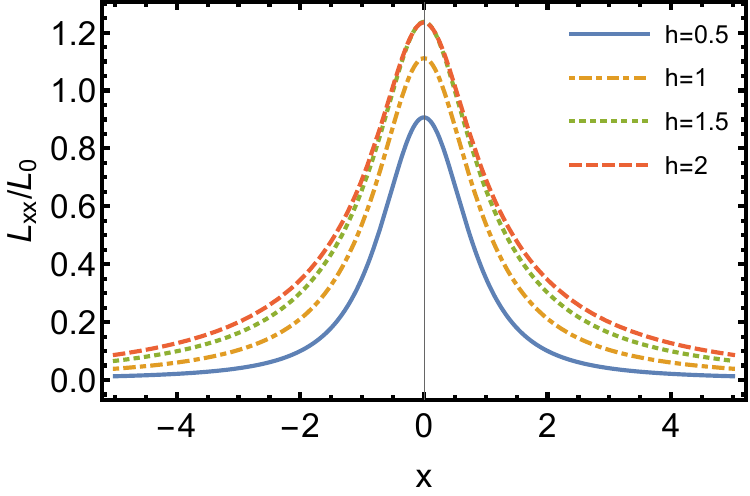}
\includegraphics[width=0.325\linewidth]{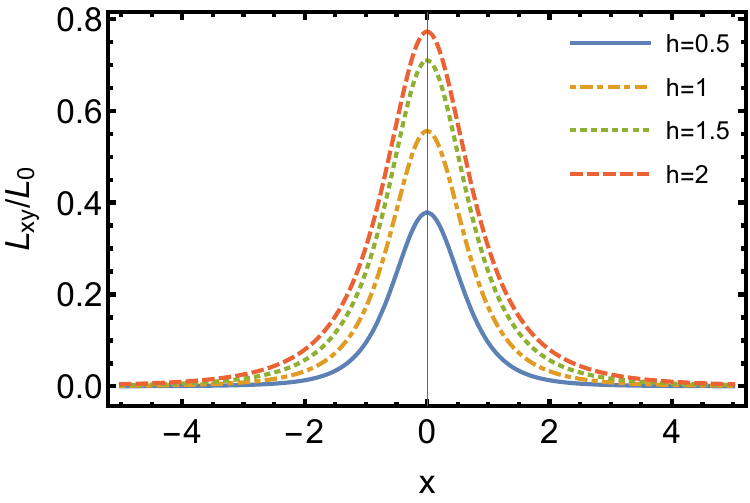}
\includegraphics[width=0.325\linewidth]{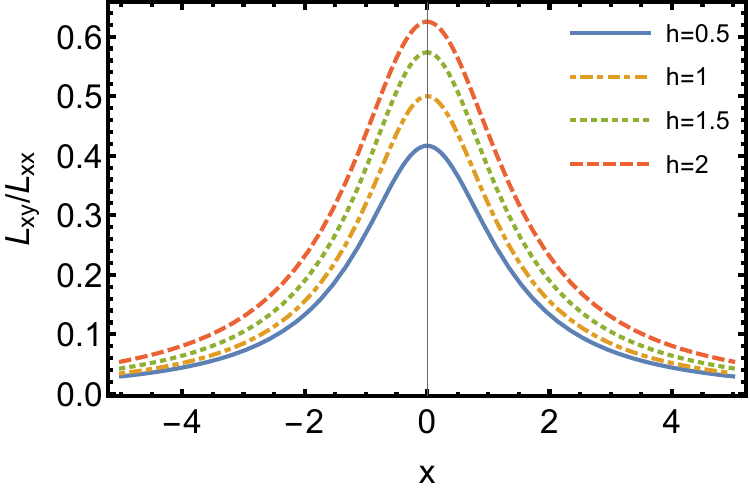}
\includegraphics[width=0.325\linewidth]{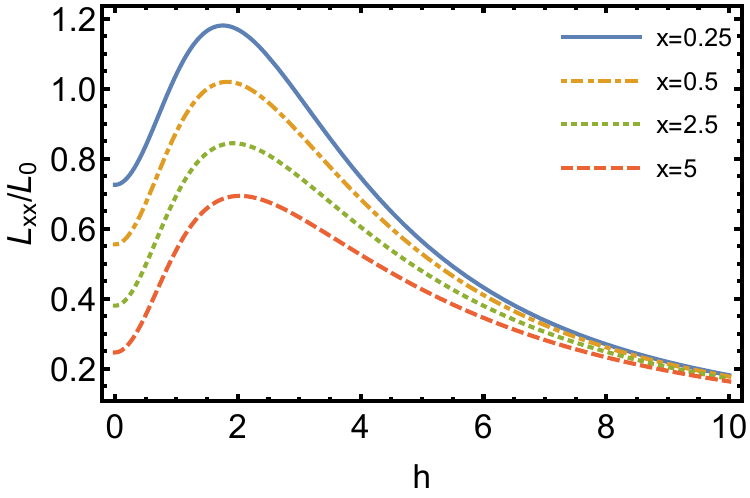}
\includegraphics[width=0.325\linewidth]{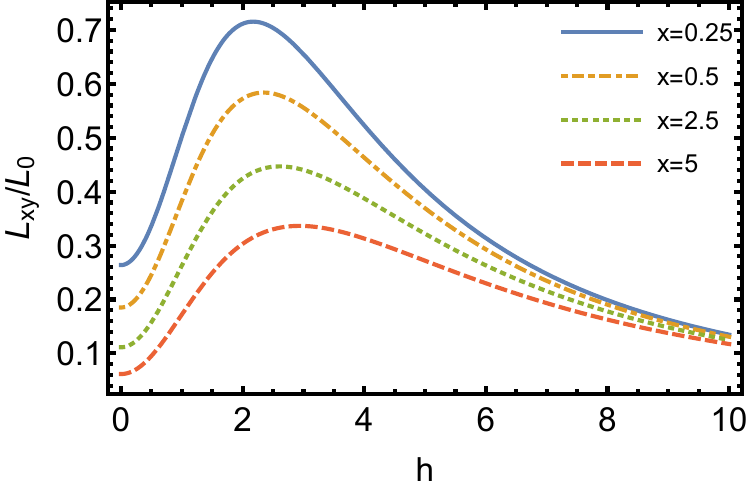}
\includegraphics[width=0.325\linewidth]{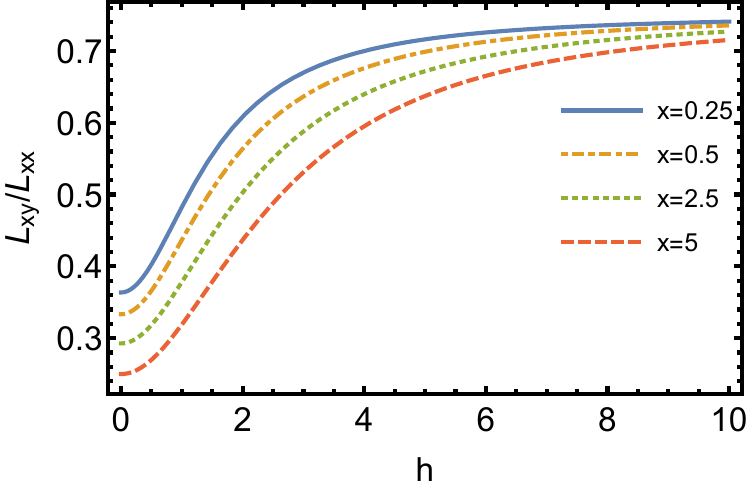}
\caption{Representative line-cuts extracted from Fig. \ref{fig:L-3D} for the density dependence of $L_{ij}(n,H)$ at several values of the magnetic field (the first three plots on the upper row) and for the field dependence at several densities (the next three plots on the lower row). Units and normalizations are the same as in Fig. \ref{fig:L-3D}.}\label{fig:L-2D}
\end{figure*}

\section{Results for the Lorenz ratio}

In this section, we present results for the Lorenz ratios extracted from the effective thermoelectric matrix under isothermal transport conditions. For a more concise presentation, it is convenient to work with dimensionless variables for both density and magnetic field. Accordingly, we introduce the following variables:
\begin{equation}\label{eq:x-h}
x=\frac{\sqrt{2}n}{\sqrt{\langle\delta n^2\rangle}}, \quad h=\frac{\sigma_0}{e^2}\frac{\sqrt{2}}{l^2_H\sqrt{\langle\delta n^2\rangle}}.
\end{equation} 
Both the particle density and the characteristic field strength are measured in units of the residual density variation, $\delta n$, induced by the disorder potential. We begin with the zero-field limit, which has been extensively studied in the literature. Revisiting this case remains useful, as different approaches have been employed in previous works. 

For the Lorenz ratio of diagonal conductivities at vanishing field, $h\to0$, we find from Eq. \eqref{eq:J}
\begin{equation}\label{eq:Lxx-x}
L_{xx}=L_0\frac{1+\chi_\sigma}{(1+\chi_\sigma+x^2)^2},\quad L_0=\frac{s^2}{\sigma_0k}, 
\end{equation} 
where $\chi_\sigma=e^2\chi/\sigma_0$ and we assumed that $Ts^2/k\gg\kappa$. This expression shows $L_{xx}$ is peaked at the charge neutrality and behaves as a square of Lorentzian at finite density. By comparing the peak height to the value of the Lorenz ratio corresponding to the WF law we can deduce  
\begin{equation}
\frac{L_0}{L_{\text{WF}}}\simeq\frac{e^2}{\sigma_0}\frac{s^2}{k}\sim \frac{\sigma_0}{\sigma_0+e^2\chi}\frac{s^2}{\langle\delta n^2\rangle},
\end{equation}
where for brevity we omitted all the numerical factors. Therefore, the peak is primarily dominated by the value of residual disorder in density fluctuations around charge neutrality. Depending on the sample quality $\delta n$ can be estimated from the low-temperature resistivity, which is disorder limited. The typical number for sufficiently clean samples is in the range $\sim 5\times 10^9$ cm$^{-2}$. The temperature dependence of the entropy density for Dirac fermions in graphene can be estimates as $s\sim (T/v)^2$. For the temperature $T\sim 100$ K and band velocity $v\sim 10^6$ m/s it gives $s\sim 10^{10}$ cm$^{-2}$. For these values $L_0/L_{\text{WF}}\sim 4$ if one takes $\sigma/\sigma_0\sim1$. This estimate sensitively depends on temperature since it scales at $\propto T^4$ in this regime.  For example, at $T\sim150$ K one finds $L_0/L_{\text{WF}}\sim 20$. At higher temperatures, the hydrodynamic approximation is believed to break down as phonon scattering becomes significant. However, even within the hydrodynamic regime, making precise predictions for the temperature dependence remains challenging. This difficulty arises because multiple quantities depend on $T$, and their behavior is not well understood theoretically beyond perturbative estimates. In part for this reason we focus on the density and field dependence. 

Another limit that allows for a simple analytical result is transport at charge neutrality in a finite magnetic field. In this case, the diagonal Lorenz ratio takes the form   
\begin{equation}\label{eq:Lxx-h}
L_{xx}=L_0\frac{(1+\chi_\sigma)(h^2+1)}{(1+\chi_\sigma(h^2+1))^2}.
\end{equation} 
We observe that a strong magnetic field tends to suppress the Lorenz ratio; however, the overall dependence is nonmonotonic, as it initially increases. This suggests that, at first, the effect of the field on magnetoresistance is stronger than its effect on thermal resistance.

In complete analogy, exploring the same limiting cases, one can extract Lorenz ratios for the Hall conductivities. At small fields the result is 
\begin{equation}\label{eq:Lxy-x}
L_{xy}=L_0\frac{1-\chi^2_\sigma}{(1+\chi_\sigma+x^2)^2(x^2+2)}.
\end{equation}
As compared to $L_{xx}$ in Eq. \eqref{eq:Lxx-x} this expression exhibits a much stronger decay at high doping. Conversely, at charge neutrality and arbitrary field one finds 
\begin{equation}\label{eq:Lxy-h}
L_{xy}=L_0\frac{(1-\chi^2_\sigma)(h^2+1)^2}{(1+\chi_\sigma(h^2+1))^2(h^2+2)}.
\end{equation}

To provide a comprehensive picture, we present a surface plot of $L_{xx}$, $L_{xy}$ and their ratio $L_{xy}/L_{xx}$ as functions of both density and magnetic field in Fig. \ref{fig:L-3D}, covering a broad range of parameters. Throughout this range, we observe that 
$L_{xy}/L_{xx}<1$, indicating that the violation of the WF law is more pronounced in the diagonal conductivities. Representative line cuts of $L_{ij}(n,H)$ as a function of density for several values of the magnetic field, as well as vice versa, are shown in Fig. \ref{fig:L-2D}. One of the most interesting features of these results is the apparent nonmonotonicity of the Lorenz ratios as a function of the magnetic field, along with an enhancement of the peak values for both $L_{xx}$ and $L_{xy}$ at intermediate field strengths. 

\section{Summary and Discussion}

In light of the pronounced thermoelectric anomalies near charge neutrality in graphene devices, it is of interest to examine their thermoelectric efficiency. This property can be quantified by the dimensionless figure of merit, defined as \cite{Mahan:1997}
\begin{equation}
ZT=\frac{T\sigma }{\kappa}Q^2
\end{equation}
where $Q$ is the thermopower (Seebeck coefficient). By the Onsager symmetry it is related to the thermoelectric Peltier coefficient $\Pi=TQ$. 
The figure of merit quantifies the efficiency of thermoelectric cooling. A divergent value of the figure of merit implies that the device approaches Carnot efficiency. 
Materials regarded as good thermoelectric coolers must exhibit high values of $ZT$; typically, $ZT>1$ is considered necessary for practical applications.

For pristine electron fluids, the Seebeck coefficient is given by the entropy per unit charge transported by the current. In the hydrodynamic limit, at relatively high doping, it approaches the universal value 
$Q=s/en$, but it always remains below this bound. The reduction is especially pronounced near charge neutrality, where we find (at zero magnetic field):
\begin{equation}
Q=\frac{2}{e}\frac{ns}{\langle\delta n^2\rangle}\frac{1}{1+\chi_\sigma+x^2}.
\end{equation}
Vanishing of $Q$ as $n\to0$ is the consequence of the particle-hole symmetry. Combining this result with the expression obtained for the Lorenz ratio in the same limit, we deduce the figure of merit in the form:
\begin{equation}
ZT=\frac{Q^2}{L}=\frac{2n^2}{\langle\delta n^2\rangle}\frac{\sigma_0}{\sigma_0+e^2\chi}.
\end{equation}
As a result, in a broad range of densities near charge neutrality, $\delta n\ll n \ll s$, it is therefore possible to have $ZT>1$.  We notice that the maximal enhancement is of the same order as that of the Lorenz ratio.

In summary, we studied the magnetotransport properties of electron liquids in monolayer graphene, focusing on the apparent violation of the Wiedemann-Franz law observed in recent experiments \cite{Talanov:2024}. Our theoretical framework applies to the hydrodynamic transport regime and qualitatively captures several key experimental features, particularly the density dependence of the Lorenz ratio near the charge neutrality. Additionally, our model provides predictions for the field dependence of the Lorenz ratio, which can be further tested in experiments. Importantly, this theory is not limited to graphene but applies more broadly to weakly disordered, non-Galilean invariant conductors. It can also be extended to incorporate additional effects, such as Hall viscosity, and adapted to various finite-size geometries, including Hall-bar and Corbino devices. The focus of the forthcoming work will be on the adiabatic regime of thermomagnetic transport.
In closing, we note that similar anomalies can arise in electron liquids with a three-dimensional Dirac spectrum, such as Weyl semimetals, as recently investigated in Ref. \cite{Messica:2023}.

\section*{Acknowledgments}

I am grateful to Anton Andreev, Philip Kim, Songci Li, Dmitrii Maslov, and Jonah Waissman for numerous useful discussions. I appreciate input from Kamran Behnia and Erez Berg for bringing relevant literature to my attention. 
I thank Ilya Esterlis for reading and commenting on the paper. This work was supported by the National Science Foundation Grant No. DMR-2452658 and H. I. Romnes Faculty Fellowship provided by the University of Wisconsin-Madison Office of the Vice Chancellor for Research and Graduate Education with funding from the Wisconsin Alumni Research Foundation.

\bibliography{biblio}

\begin{thebibliography}{53}%
\makeatletter
\providecommand \@ifxundefined [1]{%
 \@ifx{#1\undefined}
}%
\providecommand \@ifnum [1]{%
 \ifnum #1\expandafter \@firstoftwo
 \else \expandafter \@secondoftwo
 \fi
}%
\providecommand \@ifx [1]{%
 \ifx #1\expandafter \@firstoftwo
 \else \expandafter \@secondoftwo
 \fi
}%
\providecommand \natexlab [1]{#1}%
\providecommand \enquote  [1]{``#1''}%
\providecommand \bibnamefont  [1]{#1}%
\providecommand \bibfnamefont [1]{#1}%
\providecommand \citenamefont [1]{#1}%
\providecommand \href@noop [0]{\@secondoftwo}%
\providecommand \href [0]{\begingroup \@sanitize@url \@href}%
\providecommand \@href[1]{\@@startlink{#1}\@@href}%
\providecommand \@@href[1]{\endgroup#1\@@endlink}%
\providecommand \@sanitize@url [0]{\catcode `\\12\catcode `\$12\catcode
  `\&12\catcode `\#12\catcode `\^12\catcode `\_12\catcode `\%12\relax}%
\providecommand \@@startlink[1]{}%
\providecommand \@@endlink[0]{}%
\providecommand \url  [0]{\begingroup\@sanitize@url \@url }%
\providecommand \@url [1]{\endgroup\@href {#1}{\urlprefix }}%
\providecommand \urlprefix  [0]{URL }%
\providecommand \Eprint [0]{\href }%
\providecommand \doibase [0]{http://dx.doi.org/}%
\providecommand \selectlanguage [0]{\@gobble}%
\providecommand \bibinfo  [0]{\@secondoftwo}%
\providecommand \bibfield  [0]{\@secondoftwo}%
\providecommand \translation [1]{[#1]}%
\providecommand \BibitemOpen [0]{}%
\providecommand \bibitemStop [0]{}%
\providecommand \bibitemNoStop [0]{.\EOS\space}%
\providecommand \EOS [0]{\spacefactor3000\relax}%
\providecommand \BibitemShut  [1]{\csname bibitem#1\endcsname}%
\let\auto@bib@innerbib\@empty
\bibitem [{\citenamefont {Azbel}\ \emph {et~al.}(1956)\citenamefont {Azbel},
  \citenamefont {Kaganov},\ and\ \citenamefont {Lifshitz}}]{AKL:1956}%
  \BibitemOpen
  \bibfield  {author} {\bibinfo {author} {\bibfnamefont {M.~Ia.}\ \bibnamefont
  {Azbel}}, \bibinfo {author} {\bibfnamefont {M.~I.}\ \bibnamefont {Kaganov}},
  \ and\ \bibinfo {author} {\bibfnamefont {I.~M.}\ \bibnamefont {Lifshitz}},\
  }\bibfield  {title} {\enquote {\bibinfo {title} {On the theory of
  galvanomagnetic effects in metals},}\ }\href@noop {} {\bibfield  {journal}
  {\bibinfo  {journal} {Sov. Phys. JETP}\ }\textbf {\bibinfo {volume} {3}},\
  \bibinfo {pages} {143} (\bibinfo {year} {1956})}\BibitemShut {NoStop}%
\bibitem [{\citenamefont {Azbel}\ \emph
  {et~al.}(1957{\natexlab{a}})\citenamefont {Azbel}, \citenamefont {Kaganov},\
  and\ \citenamefont {Lifshitz}}]{AKL:1957a}%
  \BibitemOpen
  \bibfield  {author} {\bibinfo {author} {\bibfnamefont {M.~Ia.}\ \bibnamefont
  {Azbel}}, \bibinfo {author} {\bibfnamefont {M.~I.}\ \bibnamefont {Kaganov}},
  \ and\ \bibinfo {author} {\bibfnamefont {I.~M.}\ \bibnamefont {Lifshitz}},\
  }\bibfield  {title} {\enquote {\bibinfo {title} {The theory of
  galvanomagnetic effects in metals},}\ }\href@noop {} {\bibfield  {journal}
  {\bibinfo  {journal} {Sov. Phys. JETP}\ }\textbf {\bibinfo {volume} {4}},\
  \bibinfo {pages} {41} (\bibinfo {year} {1957}{\natexlab{a}})}\BibitemShut
  {NoStop}%
\bibitem [{\citenamefont {Azbel}\ \emph
  {et~al.}(1957{\natexlab{b}})\citenamefont {Azbel}, \citenamefont {Kaganov},\
  and\ \citenamefont {Lifshitz}}]{AKL:1957b}%
  \BibitemOpen
  \bibfield  {author} {\bibinfo {author} {\bibfnamefont {M.~Ia.}\ \bibnamefont
  {Azbel}}, \bibinfo {author} {\bibfnamefont {M.~I.}\ \bibnamefont {Kaganov}},
  \ and\ \bibinfo {author} {\bibfnamefont {I.~M.}\ \bibnamefont {Lifshitz}},\
  }\bibfield  {title} {\enquote {\bibinfo {title} {Thermal conductivity and
  thermoelectric phenomena in metals in a magnetic field},}\ }\href@noop {}
  {\bibfield  {journal} {\bibinfo  {journal} {Sov. Phys. JETP}\ }\textbf
  {\bibinfo {volume} {5}},\ \bibinfo {pages} {967} (\bibinfo {year}
  {1957}{\natexlab{b}})}\BibitemShut {NoStop}%
\bibitem [{\citenamefont {Abrikosov}(1988)}]{Abrikosov}%
  \BibitemOpen
  \bibfield  {author} {\bibinfo {author} {\bibfnamefont {A.~A.}\ \bibnamefont
  {Abrikosov}},\ }\href@noop {} {\emph {\bibinfo {title} {Fundamentals of the
  Theory of Metals}}}\ (\bibinfo  {publisher} {North Holland, Amsterdam},\
  \bibinfo {year} {1988})\BibitemShut {NoStop}%
\bibitem [{\citenamefont {Ziman}(2001)}]{Ziman}%
  \BibitemOpen
  \bibfield  {author} {\bibinfo {author} {\bibfnamefont {J.}~\bibnamefont
  {Ziman}},\ }\href@noop {} {\emph {\bibinfo {title} {Electrons and Phonons}}}\
  (\bibinfo  {publisher} {Oxford University Press, Oxford},\ \bibinfo {year}
  {2001})\BibitemShut {NoStop}%
\bibitem [{\citenamefont {Pippard}(1989)}]{Pippard}%
  \BibitemOpen
  \bibfield  {author} {\bibinfo {author} {\bibfnamefont {A.~B.}\ \bibnamefont
  {Pippard}},\ }\href@noop {} {\emph {\bibinfo {title} {Magnetoresistance in
  Metals}}}\ (\bibinfo  {publisher} {Cambridge University Press},\ \bibinfo
  {year} {1989})\BibitemShut {NoStop}%
\bibitem [{Note1()}]{Note1}%
  \BibitemOpen
  \bibinfo {note} {Throughout this paper, we adopt natural units where Planck's
  constant and Boltzmann's constant are set to unity, $\hbar =k_{\protect \text
  {B}}=1$.}\BibitemShut {Stop}%
\bibitem [{\citenamefont {Kumar}\ \emph {et~al.}(1993)\citenamefont {Kumar},
  \citenamefont {Prasad},\ and\ \citenamefont {Pohl}}]{Kumar:1993}%
  \BibitemOpen
  \bibfield  {author} {\bibinfo {author} {\bibfnamefont {G.~S.}\ \bibnamefont
  {Kumar}}, \bibinfo {author} {\bibfnamefont {G.}~\bibnamefont {Prasad}}, \
  and\ \bibinfo {author} {\bibfnamefont {R.~O.}\ \bibnamefont {Pohl}},\
  }\bibfield  {title} {\enquote {\bibinfo {title} {Experimental determinations
  of the {L}orenz number},}\ }\href {\doibase 10.1007/BF01154931} {\bibfield
  {journal} {\bibinfo  {journal} {Journal of Materials Science}\ }\textbf
  {\bibinfo {volume} {28}},\ \bibinfo {pages} {4261--4272} (\bibinfo {year}
  {1993})}\BibitemShut {NoStop}%
\bibitem [{\citenamefont {Gourgout}\ \emph {et~al.}(2024)\citenamefont
  {Gourgout}, \citenamefont {Marguerite}, \citenamefont {Fauqu\'e},\ and\
  \citenamefont {Behnia}}]{Gourgout:2024}%
  \BibitemOpen
  \bibfield  {author} {\bibinfo {author} {\bibfnamefont {Adrien}\ \bibnamefont
  {Gourgout}}, \bibinfo {author} {\bibfnamefont {Arthur}\ \bibnamefont
  {Marguerite}}, \bibinfo {author} {\bibfnamefont {Beno\^{\i}t}\ \bibnamefont
  {Fauqu\'e}}, \ and\ \bibinfo {author} {\bibfnamefont {Kamran}\ \bibnamefont
  {Behnia}},\ }\bibfield  {title} {\enquote {\bibinfo {title} {Electronic
  thermal resistivity and quasiparticle collision cross section in
  semimetals},}\ }\href {\doibase 10.1103/PhysRevB.110.155119} {\bibfield
  {journal} {\bibinfo  {journal} {Phys. Rev. B}\ }\textbf {\bibinfo {volume}
  {110}},\ \bibinfo {pages} {155119} (\bibinfo {year} {2024})}\BibitemShut
  {NoStop}%
\bibitem [{\citenamefont {Zhang}\ \emph {et~al.}(2000)\citenamefont {Zhang},
  \citenamefont {Ong}, \citenamefont {Xu}, \citenamefont {Krishana},
  \citenamefont {Gagnon},\ and\ \citenamefont {Taillefer}}]{Taillefer:2000}%
  \BibitemOpen
  \bibfield  {author} {\bibinfo {author} {\bibfnamefont {Y.}~\bibnamefont
  {Zhang}}, \bibinfo {author} {\bibfnamefont {N.~P.}\ \bibnamefont {Ong}},
  \bibinfo {author} {\bibfnamefont {Z.~A.}\ \bibnamefont {Xu}}, \bibinfo
  {author} {\bibfnamefont {K.}~\bibnamefont {Krishana}}, \bibinfo {author}
  {\bibfnamefont {R.}~\bibnamefont {Gagnon}}, \ and\ \bibinfo {author}
  {\bibfnamefont {L.}~\bibnamefont {Taillefer}},\ }\bibfield  {title} {\enquote
  {\bibinfo {title} {Determining the {W}iedemann-{F}ranz ratio from the thermal
  {H}all conductivity: Application to {C}u and
  {YB}a$_{2}${C}u$_{3}${O}$_{6.95}$},}\ }\href {\doibase
  10.1103/PhysRevLett.84.2219} {\bibfield  {journal} {\bibinfo  {journal}
  {Phys. Rev. Lett.}\ }\textbf {\bibinfo {volume} {84}},\ \bibinfo {pages}
  {2219--2222} (\bibinfo {year} {2000})}\BibitemShut {NoStop}%
\bibitem [{\citenamefont {Tanatar}\ \emph {et~al.}(2007)\citenamefont
  {Tanatar}, \citenamefont {Paglione}, \citenamefont {Petrovic},\ and\
  \citenamefont {Taillefer}}]{Tanatar:2007}%
  \BibitemOpen
  \bibfield  {author} {\bibinfo {author} {\bibfnamefont {Makariy~A.}\
  \bibnamefont {Tanatar}}, \bibinfo {author} {\bibfnamefont {Johnpierre}\
  \bibnamefont {Paglione}}, \bibinfo {author} {\bibfnamefont {Cedomir}\
  \bibnamefont {Petrovic}}, \ and\ \bibinfo {author} {\bibfnamefont {Louis}\
  \bibnamefont {Taillefer}},\ }\bibfield  {title} {\enquote {\bibinfo {title}
  {Anisotropic violation of the {W}iedemann-{F}ranz law at a quantum critical
  point},}\ }\href {\doibase 10.1126/science.1140762} {\bibfield  {journal}
  {\bibinfo  {journal} {Science}\ }\textbf {\bibinfo {volume} {316}},\ \bibinfo
  {pages} {1320--1322} (\bibinfo {year} {2007})}\BibitemShut {NoStop}%
\bibitem [{\citenamefont {Grissonnanche}\ \emph {et~al.}(2016)\citenamefont
  {Grissonnanche}, \citenamefont {Lalibert\'e}, \citenamefont
  {Dufour-Beaus\'ejour}, \citenamefont {Matusiak}, \citenamefont {Badoux},
  \citenamefont {Tafti}, \citenamefont {Michon}, \citenamefont {Riopel},
  \citenamefont {Cyr-Choini\`ere}, \citenamefont {Baglo}, \citenamefont
  {Ramshaw}, \citenamefont {Liang}, \citenamefont {Bonn}, \citenamefont
  {Hardy}, \citenamefont {Kr\"amer}, \citenamefont {LeBoeuf}, \citenamefont
  {Graf}, \citenamefont {Doiron-Leyraud},\ and\ \citenamefont
  {Taillefer}}]{Taillefer:2016}%
  \BibitemOpen
  \bibfield  {author} {\bibinfo {author} {\bibfnamefont {G.}~\bibnamefont
  {Grissonnanche}}, \bibinfo {author} {\bibfnamefont {F.}~\bibnamefont
  {Lalibert\'e}}, \bibinfo {author} {\bibfnamefont {S.}~\bibnamefont
  {Dufour-Beaus\'ejour}}, \bibinfo {author} {\bibfnamefont {M.}~\bibnamefont
  {Matusiak}}, \bibinfo {author} {\bibfnamefont {S.}~\bibnamefont {Badoux}},
  \bibinfo {author} {\bibfnamefont {F.~F.}\ \bibnamefont {Tafti}}, \bibinfo
  {author} {\bibfnamefont {B.}~\bibnamefont {Michon}}, \bibinfo {author}
  {\bibfnamefont {A.}~\bibnamefont {Riopel}}, \bibinfo {author} {\bibfnamefont
  {O.}~\bibnamefont {Cyr-Choini\`ere}}, \bibinfo {author} {\bibfnamefont
  {J.~C.}\ \bibnamefont {Baglo}}, \bibinfo {author} {\bibfnamefont {B.~J.}\
  \bibnamefont {Ramshaw}}, \bibinfo {author} {\bibfnamefont {R.}~\bibnamefont
  {Liang}}, \bibinfo {author} {\bibfnamefont {D.~A.}\ \bibnamefont {Bonn}},
  \bibinfo {author} {\bibfnamefont {W.~N.}\ \bibnamefont {Hardy}}, \bibinfo
  {author} {\bibfnamefont {S.}~\bibnamefont {Kr\"amer}}, \bibinfo {author}
  {\bibfnamefont {D.}~\bibnamefont {LeBoeuf}}, \bibinfo {author} {\bibfnamefont
  {D.}~\bibnamefont {Graf}}, \bibinfo {author} {\bibfnamefont {N.}~\bibnamefont
  {Doiron-Leyraud}}, \ and\ \bibinfo {author} {\bibfnamefont {Louis}\
  \bibnamefont {Taillefer}},\ }\bibfield  {title} {\enquote {\bibinfo {title}
  {{W}iedemann-{F}ranz law in the underdoped cuprate superconductor
  {YB}a$_{2}${C}u$_{3}${O}$_y$},}\ }\href {\doibase 10.1103/PhysRevB.93.064513}
  {\bibfield  {journal} {\bibinfo  {journal} {Phys. Rev. B}\ }\textbf {\bibinfo
  {volume} {93}},\ \bibinfo {pages} {064513} (\bibinfo {year}
  {2016})}\BibitemShut {NoStop}%
\bibitem [{\citenamefont {Jaoui}\ \emph {et~al.}(2018)\citenamefont {Jaoui},
  \citenamefont {Fauqu\'e}, \citenamefont {Rischau}, \citenamefont {Subedi},
  \citenamefont {Fu}, \citenamefont {Gooth}, \citenamefont {Kumar},
  \citenamefont {S{\"u}{\ss}}, \citenamefont {Maslov}, \citenamefont {Felser},\
  and\ \citenamefont {Behnia}}]{Jaoui:2018}%
  \BibitemOpen
  \bibfield  {author} {\bibinfo {author} {\bibfnamefont {Alexandre}\
  \bibnamefont {Jaoui}}, \bibinfo {author} {\bibfnamefont {Beno\^{\i}t}\
  \bibnamefont {Fauqu\'e}}, \bibinfo {author} {\bibfnamefont {Carl~Willem}\
  \bibnamefont {Rischau}}, \bibinfo {author} {\bibfnamefont {Alaska}\
  \bibnamefont {Subedi}}, \bibinfo {author} {\bibfnamefont {Chenguang}\
  \bibnamefont {Fu}}, \bibinfo {author} {\bibfnamefont {Johannes}\ \bibnamefont
  {Gooth}}, \bibinfo {author} {\bibfnamefont {Nitesh}\ \bibnamefont {Kumar}},
  \bibinfo {author} {\bibfnamefont {Vicky}\ \bibnamefont {S{\"u}{\ss}}},
  \bibinfo {author} {\bibfnamefont {Dmitrii~L.}\ \bibnamefont {Maslov}},
  \bibinfo {author} {\bibfnamefont {Claudia}\ \bibnamefont {Felser}}, \ and\
  \bibinfo {author} {\bibfnamefont {Kamran}\ \bibnamefont {Behnia}},\
  }\bibfield  {title} {\enquote {\bibinfo {title} {Departure from the
  {W}iedemann--{F}ranz law in {WP}$_2$ driven by mismatch in {T}-square
  resistivity prefactors},}\ }\href {\doibase 10.1038/s41535-018-0136-x}
  {\bibfield  {journal} {\bibinfo  {journal} {npj Quantum Materials}\ }\textbf
  {\bibinfo {volume} {3}},\ \bibinfo {pages} {64} (\bibinfo {year}
  {2018})}\BibitemShut {NoStop}%
\bibitem [{\citenamefont {Gooth}\ \emph {et~al.}(2018)\citenamefont {Gooth},
  \citenamefont {Menges}, \citenamefont {Kumar}, \citenamefont {S{\"u}{\ss}},
  \citenamefont {Shekhar}, \citenamefont {Sun}, \citenamefont {Drechsler},
  \citenamefont {Zierold}, \citenamefont {Felser},\ and\ \citenamefont
  {Gotsmann}}]{Gooth:2018}%
  \BibitemOpen
  \bibfield  {author} {\bibinfo {author} {\bibfnamefont {J.}~\bibnamefont
  {Gooth}}, \bibinfo {author} {\bibfnamefont {F.}~\bibnamefont {Menges}},
  \bibinfo {author} {\bibfnamefont {N.}~\bibnamefont {Kumar}}, \bibinfo
  {author} {\bibfnamefont {V.}~\bibnamefont {S{\"u}{\ss}}}, \bibinfo {author}
  {\bibfnamefont {C.}~\bibnamefont {Shekhar}}, \bibinfo {author} {\bibfnamefont
  {Y.}~\bibnamefont {Sun}}, \bibinfo {author} {\bibfnamefont {U.}~\bibnamefont
  {Drechsler}}, \bibinfo {author} {\bibfnamefont {R.}~\bibnamefont {Zierold}},
  \bibinfo {author} {\bibfnamefont {C.}~\bibnamefont {Felser}}, \ and\ \bibinfo
  {author} {\bibfnamefont {B.}~\bibnamefont {Gotsmann}},\ }\bibfield  {title}
  {\enquote {\bibinfo {title} {Thermal and electrical signatures of a
  hydrodynamic electron fluid in tungsten diphosphide},}\ }\href {\doibase
  10.1038/s41467-018-06688-y} {\bibfield  {journal} {\bibinfo  {journal}
  {Nature Communications}\ }\textbf {\bibinfo {volume} {9}},\ \bibinfo {pages}
  {4093} (\bibinfo {year} {2018})}\BibitemShut {NoStop}%
\bibitem [{\citenamefont {Xu}\ \emph {et~al.}(2020)\citenamefont {Xu},
  \citenamefont {Li}, \citenamefont {Lu}, \citenamefont {Collignon},
  \citenamefont {Fu}, \citenamefont {Koo}, \citenamefont {Fauqu\'e},
  \citenamefont {Yan}, \citenamefont {Zhu},\ and\ \citenamefont
  {Behnia}}]{Behnia:2020}%
  \BibitemOpen
  \bibfield  {author} {\bibinfo {author} {\bibfnamefont {Liangcai}\
  \bibnamefont {Xu}}, \bibinfo {author} {\bibfnamefont {Xiaokang}\ \bibnamefont
  {Li}}, \bibinfo {author} {\bibfnamefont {Xiufang}\ \bibnamefont {Lu}},
  \bibinfo {author} {\bibfnamefont {Cl\'ement}\ \bibnamefont {Collignon}},
  \bibinfo {author} {\bibfnamefont {Huixia}\ \bibnamefont {Fu}}, \bibinfo
  {author} {\bibfnamefont {Jahyun}\ \bibnamefont {Koo}}, \bibinfo {author}
  {\bibfnamefont {Beno\^{\i}t}\ \bibnamefont {Fauqu\'e}}, \bibinfo {author}
  {\bibfnamefont {Binghai}\ \bibnamefont {Yan}}, \bibinfo {author}
  {\bibfnamefont {Zengwei}\ \bibnamefont {Zhu}}, \ and\ \bibinfo {author}
  {\bibfnamefont {Kamran}\ \bibnamefont {Behnia}},\ }\bibfield  {title}
  {\enquote {\bibinfo {title} {Finite-temperature violation of the anomalous
  transverse {W}iedemann-{F}ranz law},}\ }\href {\doibase
  10.1126/sciadv.aaz3522} {\bibfield  {journal} {\bibinfo  {journal} {Science
  Advances}\ }\textbf {\bibinfo {volume} {6}},\ \bibinfo {pages} {eaaz3522}
  (\bibinfo {year} {2020})}\BibitemShut {NoStop}%
\bibitem [{\citenamefont {Jiang}\ \emph {et~al.}(2023)\citenamefont {Jiang},
  \citenamefont {Fauqu\'e},\ and\ \citenamefont {Behnia}}]{Behnia:2023}%
  \BibitemOpen
  \bibfield  {author} {\bibinfo {author} {\bibfnamefont {Shan}\ \bibnamefont
  {Jiang}}, \bibinfo {author} {\bibfnamefont {Beno\^{\i}t}\ \bibnamefont
  {Fauqu\'e}}, \ and\ \bibinfo {author} {\bibfnamefont {Kamran}\ \bibnamefont
  {Behnia}},\ }\bibfield  {title} {\enquote {\bibinfo {title} {{T}-square
  dependence of the electronic thermal resistivity of metallic strontium
  titanate},}\ }\href {\doibase 10.1103/PhysRevLett.131.016301} {\bibfield
  {journal} {\bibinfo  {journal} {Phys. Rev. Lett.}\ }\textbf {\bibinfo
  {volume} {131}},\ \bibinfo {pages} {016301} (\bibinfo {year}
  {2023})}\BibitemShut {NoStop}%
\bibitem [{\citenamefont {Sun}\ \emph {et~al.}(2024)\citenamefont {Sun},
  \citenamefont {Mishra}, \citenamefont {Stockert}, \citenamefont {Daou},
  \citenamefont {Kikugawa}, \citenamefont {Perry}, \citenamefont {Hassinger},
  \citenamefont {Hartnoll}, \citenamefont {Mackenzie},\ and\ \citenamefont
  {Sunko}}]{Sunko:2024}%
  \BibitemOpen
  \bibfield  {author} {\bibinfo {author} {\bibfnamefont {Fei}\ \bibnamefont
  {Sun}}, \bibinfo {author} {\bibfnamefont {Simli}\ \bibnamefont {Mishra}},
  \bibinfo {author} {\bibfnamefont {Ulrike}\ \bibnamefont {Stockert}}, \bibinfo
  {author} {\bibfnamefont {Ramzy}\ \bibnamefont {Daou}}, \bibinfo {author}
  {\bibfnamefont {Naoki}\ \bibnamefont {Kikugawa}}, \bibinfo {author}
  {\bibfnamefont {Robin~S.}\ \bibnamefont {Perry}}, \bibinfo {author}
  {\bibfnamefont {Elena}\ \bibnamefont {Hassinger}}, \bibinfo {author}
  {\bibfnamefont {Sean~A.}\ \bibnamefont {Hartnoll}}, \bibinfo {author}
  {\bibfnamefont {Andrew~P.}\ \bibnamefont {Mackenzie}}, \ and\ \bibinfo
  {author} {\bibfnamefont {Veronika}\ \bibnamefont {Sunko}},\ }\bibfield
  {title} {\enquote {\bibinfo {title} {The {L}orenz ratio as a guide to
  scattering contributions to transport in strongly correlated metals},}\
  }\href {\doibase 10.1073/pnas.2318159121} {\bibfield  {journal} {\bibinfo
  {journal} {Proceedings of the National Academy of Sciences}\ }\textbf
  {\bibinfo {volume} {121}},\ \bibinfo {pages} {e2318159121} (\bibinfo {year}
  {2024})}\BibitemShut {NoStop}%
\bibitem [{\citenamefont {Li}(2002)}]{Li:2002}%
  \BibitemOpen
  \bibfield  {author} {\bibinfo {author} {\bibfnamefont {Mei-Rong}\
  \bibnamefont {Li}},\ }\bibfield  {title} {\enquote {\bibinfo {title} {Thermal
  {H}all conductivity of marginal {F}ermi liquids subject to out-of-plane
  impurities in high-{T}$_{c}$ cuprates},}\ }\href {\doibase
  10.1103/PhysRevB.65.184515} {\bibfield  {journal} {\bibinfo  {journal} {Phys.
  Rev. B}\ }\textbf {\bibinfo {volume} {65}},\ \bibinfo {pages} {184515}
  (\bibinfo {year} {2002})}\BibitemShut {NoStop}%
\bibitem [{\citenamefont {Mahajan}\ \emph {et~al.}(2013)\citenamefont
  {Mahajan}, \citenamefont {Barkeshli},\ and\ \citenamefont
  {Hartnoll}}]{Mahajan:2013}%
  \BibitemOpen
  \bibfield  {author} {\bibinfo {author} {\bibfnamefont {Raghu}\ \bibnamefont
  {Mahajan}}, \bibinfo {author} {\bibfnamefont {Maissam}\ \bibnamefont
  {Barkeshli}}, \ and\ \bibinfo {author} {\bibfnamefont {Sean~A.}\ \bibnamefont
  {Hartnoll}},\ }\bibfield  {title} {\enquote {\bibinfo {title} {Non-{F}ermi
  liquids and the {W}iedemann-{F}ranz law},}\ }\href {\doibase
  10.1103/PhysRevB.88.125107} {\bibfield  {journal} {\bibinfo  {journal} {Phys.
  Rev. B}\ }\textbf {\bibinfo {volume} {88}},\ \bibinfo {pages} {125107}
  (\bibinfo {year} {2013})}\BibitemShut {NoStop}%
\bibitem [{\citenamefont {Song}\ \emph {et~al.}(2017)\citenamefont {Song},
  \citenamefont {Jian},\ and\ \citenamefont {Balents}}]{Balents:2017}%
  \BibitemOpen
  \bibfield  {author} {\bibinfo {author} {\bibfnamefont {Xue-Yang}\
  \bibnamefont {Song}}, \bibinfo {author} {\bibfnamefont {Chao-Ming}\
  \bibnamefont {Jian}}, \ and\ \bibinfo {author} {\bibfnamefont {Leon}\
  \bibnamefont {Balents}},\ }\bibfield  {title} {\enquote {\bibinfo {title}
  {Strongly correlated metal built from {S}achdev-{Y}e-{K}itaev models},}\
  }\href {\doibase 10.1103/PhysRevLett.119.216601} {\bibfield  {journal}
  {\bibinfo  {journal} {Phys. Rev. Lett.}\ }\textbf {\bibinfo {volume} {119}},\
  \bibinfo {pages} {216601} (\bibinfo {year} {2017})}\BibitemShut {NoStop}%
\bibitem [{\citenamefont {Li}\ and\ \citenamefont {Maslov}(2018)}]{Li:2018}%
  \BibitemOpen
  \bibfield  {author} {\bibinfo {author} {\bibfnamefont {Songci}\ \bibnamefont
  {Li}}\ and\ \bibinfo {author} {\bibfnamefont {Dmitrii~L.}\ \bibnamefont
  {Maslov}},\ }\bibfield  {title} {\enquote {\bibinfo {title} {Lorentz ratio of
  a compensated metal},}\ }\href {\doibase 10.1103/PhysRevB.98.245134}
  {\bibfield  {journal} {\bibinfo  {journal} {Phys. Rev. B}\ }\textbf {\bibinfo
  {volume} {98}},\ \bibinfo {pages} {245134} (\bibinfo {year}
  {2018})}\BibitemShut {NoStop}%
\bibitem [{\citenamefont {Levchenko}\ and\ \citenamefont
  {Schmalian}(2020)}]{Levchenko:2020}%
  \BibitemOpen
  \bibfield  {author} {\bibinfo {author} {\bibfnamefont {Alex}\ \bibnamefont
  {Levchenko}}\ and\ \bibinfo {author} {\bibfnamefont {J\"org}\ \bibnamefont
  {Schmalian}},\ }\bibfield  {title} {\enquote {\bibinfo {title} {Transport
  properties of strongly coupled electron--phonon liquids},}\ }\href {\doibase
  https://doi.org/10.1016/j.aop.2020.168218} {\bibfield  {journal} {\bibinfo
  {journal} {Annals of Physics}\ }\textbf {\bibinfo {volume} {419}},\ \bibinfo
  {pages} {168218} (\bibinfo {year} {2020})}\BibitemShut {NoStop}%
\bibitem [{\citenamefont {Huang}\ and\ \citenamefont
  {Lucas}(2021)}]{Huang:2021}%
  \BibitemOpen
  \bibfield  {author} {\bibinfo {author} {\bibfnamefont {Xiaoyang}\
  \bibnamefont {Huang}}\ and\ \bibinfo {author} {\bibfnamefont {Andrew}\
  \bibnamefont {Lucas}},\ }\bibfield  {title} {\enquote {\bibinfo {title}
  {Electron-phonon hydrodynamics},}\ }\href {\doibase
  10.1103/PhysRevB.103.155128} {\bibfield  {journal} {\bibinfo  {journal}
  {Phys. Rev. B}\ }\textbf {\bibinfo {volume} {103}},\ \bibinfo {pages}
  {155128} (\bibinfo {year} {2021})}\BibitemShut {NoStop}%
\bibitem [{\citenamefont {Tulipman}\ and\ \citenamefont
  {Berg}(2023)}]{Tulipman:2023}%
  \BibitemOpen
  \bibfield  {author} {\bibinfo {author} {\bibfnamefont {Evyatar}\ \bibnamefont
  {Tulipman}}\ and\ \bibinfo {author} {\bibfnamefont {Erez}\ \bibnamefont
  {Berg}},\ }\bibfield  {title} {\enquote {\bibinfo {title} {A criterion for
  strange metallicity in the {L}orenz ratio},}\ }\href {\doibase
  10.1038/s41535-023-00598-z} {\bibfield  {journal} {\bibinfo  {journal} {npj
  Quantum Materials}\ }\textbf {\bibinfo {volume} {8}},\ \bibinfo {pages} {66}
  (\bibinfo {year} {2023})}\BibitemShut {NoStop}%
\bibitem [{\citenamefont {Gurzhi}(1968)}]{Gurzhi:1968}%
  \BibitemOpen
  \bibfield  {author} {\bibinfo {author} {\bibfnamefont {R.~N.}\ \bibnamefont
  {Gurzhi}},\ }\bibfield  {title} {\enquote {\bibinfo {title} {Hydrodynamic
  effects in solids at low temperature},}\ }\href@noop {} {\bibfield  {journal}
  {\bibinfo  {journal} {Soviet Physics Uspekhi}\ }\textbf {\bibinfo {volume}
  {11}},\ \bibinfo {pages} {255} (\bibinfo {year} {1968})}\BibitemShut
  {NoStop}%
\bibitem [{\citenamefont {Crossno}\ \emph {et~al.}(2016)\citenamefont
  {Crossno}, \citenamefont {Shi}, \citenamefont {Wang}, \citenamefont {Liu},
  \citenamefont {Harzheim}, \citenamefont {Lucas}, \citenamefont {Sachdev},
  \citenamefont {Kim}, \citenamefont {Taniguchi}, \citenamefont {Watanabe},
  \citenamefont {Ohki},\ and\ \citenamefont {Fong}}]{Crossno:2016}%
  \BibitemOpen
  \bibfield  {author} {\bibinfo {author} {\bibfnamefont {Jesse}\ \bibnamefont
  {Crossno}}, \bibinfo {author} {\bibfnamefont {Jing~K.}\ \bibnamefont {Shi}},
  \bibinfo {author} {\bibfnamefont {Ke}~\bibnamefont {Wang}}, \bibinfo {author}
  {\bibfnamefont {Xiaomeng}\ \bibnamefont {Liu}}, \bibinfo {author}
  {\bibfnamefont {Achim}\ \bibnamefont {Harzheim}}, \bibinfo {author}
  {\bibfnamefont {Andrew}\ \bibnamefont {Lucas}}, \bibinfo {author}
  {\bibfnamefont {Subir}\ \bibnamefont {Sachdev}}, \bibinfo {author}
  {\bibfnamefont {Philip}\ \bibnamefont {Kim}}, \bibinfo {author}
  {\bibfnamefont {Takashi}\ \bibnamefont {Taniguchi}}, \bibinfo {author}
  {\bibfnamefont {Kenji}\ \bibnamefont {Watanabe}}, \bibinfo {author}
  {\bibfnamefont {Thomas~A.}\ \bibnamefont {Ohki}}, \ and\ \bibinfo {author}
  {\bibfnamefont {Kin~Chung}\ \bibnamefont {Fong}},\ }\bibfield  {title}
  {\enquote {\bibinfo {title} {Observation of the {D}irac fluid and the
  breakdown of the {W}iedemann-{F}ranz law in graphene},}\ }\href {\doibase
  10.1126/science.aad0343} {\bibfield  {journal} {\bibinfo  {journal}
  {Science}\ }\textbf {\bibinfo {volume} {351}},\ \bibinfo {pages} {1058--1061}
  (\bibinfo {year} {2016})}\BibitemShut {NoStop}%
\bibitem [{\citenamefont {Principi}\ and\ \citenamefont
  {Vignale}(2015)}]{Principi:2015}%
  \BibitemOpen
  \bibfield  {author} {\bibinfo {author} {\bibfnamefont {Alessandro}\
  \bibnamefont {Principi}}\ and\ \bibinfo {author} {\bibfnamefont {Giovanni}\
  \bibnamefont {Vignale}},\ }\bibfield  {title} {\enquote {\bibinfo {title}
  {Violation of the {W}iedemann-{F}ranz law in hydrodynamic electron
  liquids},}\ }\href {\doibase 10.1103/PhysRevLett.115.056603} {\bibfield
  {journal} {\bibinfo  {journal} {Phys. Rev. Lett.}\ }\textbf {\bibinfo
  {volume} {115}},\ \bibinfo {pages} {056603} (\bibinfo {year}
  {2015})}\BibitemShut {NoStop}%
\bibitem [{\citenamefont {Lucas}\ \emph {et~al.}(2016)\citenamefont {Lucas},
  \citenamefont {Crossno}, \citenamefont {Fong}, \citenamefont {Kim},\ and\
  \citenamefont {Sachdev}}]{Lucas:2016}%
  \BibitemOpen
  \bibfield  {author} {\bibinfo {author} {\bibfnamefont {Andrew}\ \bibnamefont
  {Lucas}}, \bibinfo {author} {\bibfnamefont {Jesse}\ \bibnamefont {Crossno}},
  \bibinfo {author} {\bibfnamefont {Kin~Chung}\ \bibnamefont {Fong}}, \bibinfo
  {author} {\bibfnamefont {Philip}\ \bibnamefont {Kim}}, \ and\ \bibinfo
  {author} {\bibfnamefont {Subir}\ \bibnamefont {Sachdev}},\ }\bibfield
  {title} {\enquote {\bibinfo {title} {Transport in inhomogeneous quantum
  critical fluids and in the {D}irac fluid in graphene},}\ }\href {\doibase
  10.1103/PhysRevB.93.075426} {\bibfield  {journal} {\bibinfo  {journal} {Phys.
  Rev. B}\ }\textbf {\bibinfo {volume} {93}},\ \bibinfo {pages} {075426}
  (\bibinfo {year} {2016})}\BibitemShut {NoStop}%
\bibitem [{\citenamefont {Xie}\ and\ \citenamefont
  {Foster}(2016)}]{Foster:2016}%
  \BibitemOpen
  \bibfield  {author} {\bibinfo {author} {\bibfnamefont {Hong-Yi}\ \bibnamefont
  {Xie}}\ and\ \bibinfo {author} {\bibfnamefont {Matthew~S.}\ \bibnamefont
  {Foster}},\ }\bibfield  {title} {\enquote {\bibinfo {title} {Transport
  coefficients of graphene: Interplay of impurity scattering, {C}oulomb
  interaction, and optical phonons},}\ }\href {\doibase
  10.1103/PhysRevB.93.195103} {\bibfield  {journal} {\bibinfo  {journal} {Phys.
  Rev. B}\ }\textbf {\bibinfo {volume} {93}},\ \bibinfo {pages} {195103}
  (\bibinfo {year} {2016})}\BibitemShut {NoStop}%
\bibitem [{\citenamefont {Lucas}\ and\ \citenamefont
  {Das~Sarma}(2018)}]{Lucas:2018}%
  \BibitemOpen
  \bibfield  {author} {\bibinfo {author} {\bibfnamefont {Andrew}\ \bibnamefont
  {Lucas}}\ and\ \bibinfo {author} {\bibfnamefont {Sankar}\ \bibnamefont
  {Das~Sarma}},\ }\bibfield  {title} {\enquote {\bibinfo {title} {Electronic
  hydrodynamics and the breakdown of the {W}iedemann-{F}ranz and {M}ott laws in
  interacting metals},}\ }\href {\doibase 10.1103/PhysRevB.97.245128}
  {\bibfield  {journal} {\bibinfo  {journal} {Phys. Rev. B}\ }\textbf {\bibinfo
  {volume} {97}},\ \bibinfo {pages} {245128} (\bibinfo {year}
  {2018})}\BibitemShut {NoStop}%
\bibitem [{\citenamefont {Zarenia}\ \emph {et~al.}(2019)\citenamefont
  {Zarenia}, \citenamefont {Smith}, \citenamefont {Principi},\ and\
  \citenamefont {Vignale}}]{Zarenia:2019}%
  \BibitemOpen
  \bibfield  {author} {\bibinfo {author} {\bibfnamefont {Mohammad}\
  \bibnamefont {Zarenia}}, \bibinfo {author} {\bibfnamefont {Thomas~Benjamin}\
  \bibnamefont {Smith}}, \bibinfo {author} {\bibfnamefont {Alessandro}\
  \bibnamefont {Principi}}, \ and\ \bibinfo {author} {\bibfnamefont {Giovanni}\
  \bibnamefont {Vignale}},\ }\bibfield  {title} {\enquote {\bibinfo {title}
  {Breakdown of the {W}iedemann-{F}ranz law in {AB}-stacked bilayer
  graphene},}\ }\href {\doibase 10.1103/PhysRevB.99.161407} {\bibfield
  {journal} {\bibinfo  {journal} {Phys. Rev. B}\ }\textbf {\bibinfo {volume}
  {99}},\ \bibinfo {pages} {161407} (\bibinfo {year} {2019})}\BibitemShut
  {NoStop}%
\bibitem [{\citenamefont {Li}\ \emph {et~al.}(2020)\citenamefont {Li},
  \citenamefont {Levchenko},\ and\ \citenamefont {Andreev}}]{Li:2020}%
  \BibitemOpen
  \bibfield  {author} {\bibinfo {author} {\bibfnamefont {Songci}\ \bibnamefont
  {Li}}, \bibinfo {author} {\bibfnamefont {Alex}\ \bibnamefont {Levchenko}}, \
  and\ \bibinfo {author} {\bibfnamefont {A.~V.}\ \bibnamefont {Andreev}},\
  }\bibfield  {title} {\enquote {\bibinfo {title} {Hydrodynamic electron
  transport near charge neutrality},}\ }\href {\doibase
  10.1103/PhysRevB.102.075305} {\bibfield  {journal} {\bibinfo  {journal}
  {Phys. Rev. B}\ }\textbf {\bibinfo {volume} {102}},\ \bibinfo {pages}
  {075305} (\bibinfo {year} {2020})}\BibitemShut {NoStop}%
\bibitem [{\citenamefont {Li}\ \emph {et~al.}(2022)\citenamefont {Li},
  \citenamefont {Andreev},\ and\ \citenamefont {Levchenko}}]{Li:2022}%
  \BibitemOpen
  \bibfield  {author} {\bibinfo {author} {\bibfnamefont {Songci}\ \bibnamefont
  {Li}}, \bibinfo {author} {\bibfnamefont {A.~V.}\ \bibnamefont {Andreev}}, \
  and\ \bibinfo {author} {\bibfnamefont {Alex}\ \bibnamefont {Levchenko}},\
  }\bibfield  {title} {\enquote {\bibinfo {title} {Hydrodynamic electron
  transport in graphene {H}all-bar devices},}\ }\href {\doibase
  10.1103/PhysRevB.105.155307} {\bibfield  {journal} {\bibinfo  {journal}
  {Phys. Rev. B}\ }\textbf {\bibinfo {volume} {105}},\ \bibinfo {pages}
  {155307} (\bibinfo {year} {2022})}\BibitemShut {NoStop}%
\bibitem [{\citenamefont {Tu}\ and\ \citenamefont
  {Das~Sarma}(2023{\natexlab{a}})}]{Tu:2023a}%
  \BibitemOpen
  \bibfield  {author} {\bibinfo {author} {\bibfnamefont {Yi-Ting}\ \bibnamefont
  {Tu}}\ and\ \bibinfo {author} {\bibfnamefont {Sankar}\ \bibnamefont
  {Das~Sarma}},\ }\bibfield  {title} {\enquote {\bibinfo {title}
  {{W}iedemann-{F}ranz law in graphene},}\ }\href {\doibase
  10.1103/PhysRevB.107.085401} {\bibfield  {journal} {\bibinfo  {journal}
  {Phys. Rev. B}\ }\textbf {\bibinfo {volume} {107}},\ \bibinfo {pages}
  {085401} (\bibinfo {year} {2023}{\natexlab{a}})}\BibitemShut {NoStop}%
\bibitem [{\citenamefont {Tu}\ and\ \citenamefont
  {Das~Sarma}(2023{\natexlab{b}})}]{Tu:2023b}%
  \BibitemOpen
  \bibfield  {author} {\bibinfo {author} {\bibfnamefont {Yi-Ting}\ \bibnamefont
  {Tu}}\ and\ \bibinfo {author} {\bibfnamefont {Sankar}\ \bibnamefont
  {Das~Sarma}},\ }\bibfield  {title} {\enquote {\bibinfo {title}
  {{W}iedemann-{F}ranz law in graphene in the presence of a weak magnetic
  field},}\ }\href {\doibase 10.1103/PhysRevB.108.245415} {\bibfield  {journal}
  {\bibinfo  {journal} {Phys. Rev. B}\ }\textbf {\bibinfo {volume} {108}},\
  \bibinfo {pages} {245415} (\bibinfo {year} {2023}{\natexlab{b}})}\BibitemShut
  {NoStop}%
\bibitem [{\citenamefont {Kashuba}(2008)}]{Kashuba:2008}%
  \BibitemOpen
  \bibfield  {author} {\bibinfo {author} {\bibfnamefont {Alexander~B.}\
  \bibnamefont {Kashuba}},\ }\bibfield  {title} {\enquote {\bibinfo {title}
  {Conductivity of defectless graphene},}\ }\href {\doibase
  10.1103/PhysRevB.78.085415} {\bibfield  {journal} {\bibinfo  {journal} {Phys.
  Rev. B}\ }\textbf {\bibinfo {volume} {78}},\ \bibinfo {pages} {085415}
  (\bibinfo {year} {2008})}\BibitemShut {NoStop}%
\bibitem [{\citenamefont {Fritz}\ \emph {et~al.}(2008)\citenamefont {Fritz},
  \citenamefont {Schmalian}, \citenamefont {M\"uller},\ and\ \citenamefont
  {Sachdev}}]{Fritz:2008}%
  \BibitemOpen
  \bibfield  {author} {\bibinfo {author} {\bibfnamefont {Lars}\ \bibnamefont
  {Fritz}}, \bibinfo {author} {\bibfnamefont {J\"org}\ \bibnamefont
  {Schmalian}}, \bibinfo {author} {\bibfnamefont {Markus}\ \bibnamefont
  {M\"uller}}, \ and\ \bibinfo {author} {\bibfnamefont {Subir}\ \bibnamefont
  {Sachdev}},\ }\bibfield  {title} {\enquote {\bibinfo {title} {Quantum
  critical transport in clean graphene},}\ }\href {\doibase
  10.1103/PhysRevB.78.085416} {\bibfield  {journal} {\bibinfo  {journal} {Phys.
  Rev. B}\ }\textbf {\bibinfo {volume} {78}},\ \bibinfo {pages} {085416}
  (\bibinfo {year} {2008})}\BibitemShut {NoStop}%
\bibitem [{\citenamefont {Zeng}\ \emph {et~al.}(2024)\citenamefont {Zeng},
  \citenamefont {Guo}, \citenamefont {Ghosh}, \citenamefont {Watanabe},
  \citenamefont {Taniguchi}, \citenamefont {Levitov},\ and\ \citenamefont
  {Dean}}]{Zeng:2024}%
  \BibitemOpen
  \bibfield  {author} {\bibinfo {author} {\bibfnamefont {Yihang}\ \bibnamefont
  {Zeng}}, \bibinfo {author} {\bibfnamefont {Haoyu}\ \bibnamefont {Guo}},
  \bibinfo {author} {\bibfnamefont {Olivia~M.}\ \bibnamefont {Ghosh}}, \bibinfo
  {author} {\bibfnamefont {Kenji}\ \bibnamefont {Watanabe}}, \bibinfo {author}
  {\bibfnamefont {Takashi}\ \bibnamefont {Taniguchi}}, \bibinfo {author}
  {\bibfnamefont {Leonid~S.}\ \bibnamefont {Levitov}}, \ and\ \bibinfo {author}
  {\bibfnamefont {Cory~R.}\ \bibnamefont {Dean}},\ }\href
  {https://arxiv.org/abs/2407.05026} {\enquote {\bibinfo {title} {Quantitative
  measurement of viscosity in two-dimensional electron fluids},}\ } (\bibinfo
  {year} {2024}),\ \Eprint {http://arxiv.org/abs/2407.05026} {arXiv:2407.05026
  [cond-mat.mes-hall]} \BibitemShut {NoStop}%
\bibitem [{\citenamefont {Talanov}\ \emph {et~al.}(2024)\citenamefont
  {Talanov}, \citenamefont {Waissman}, \citenamefont {Hui}, \citenamefont
  {Skinner}, \citenamefont {Watanabe}, \citenamefont {Taniguchi},\ and\
  \citenamefont {Kim}}]{Talanov:2024}%
  \BibitemOpen
  \bibfield  {author} {\bibinfo {author} {\bibfnamefont {Artem}\ \bibnamefont
  {Talanov}}, \bibinfo {author} {\bibfnamefont {Jonah}\ \bibnamefont
  {Waissman}}, \bibinfo {author} {\bibfnamefont {Aaron}\ \bibnamefont {Hui}},
  \bibinfo {author} {\bibfnamefont {Brian}\ \bibnamefont {Skinner}}, \bibinfo
  {author} {\bibfnamefont {Kenji}\ \bibnamefont {Watanabe}}, \bibinfo {author}
  {\bibfnamefont {Takashi}\ \bibnamefont {Taniguchi}}, \ and\ \bibinfo {author}
  {\bibfnamefont {Philip}\ \bibnamefont {Kim}},\ }\href
  {https://arxiv.org/abs/2406.13799} {\enquote {\bibinfo {title} {Observation
  of electronic viscous dissipation in graphene magneto-thermal transport},}\ }
  (\bibinfo {year} {2024}),\ \Eprint {http://arxiv.org/abs/2406.13799}
  {arXiv:2406.13799 [cond-mat.mes-hall]} \BibitemShut {NoStop}%
\bibitem [{\citenamefont {Narozhny}\ \emph {et~al.}(2015)\citenamefont
  {Narozhny}, \citenamefont {Gornyi}, \citenamefont {Titov}, \citenamefont
  {Sch\"utt},\ and\ \citenamefont {Mirlin}}]{Narozhny:2015}%
  \BibitemOpen
  \bibfield  {author} {\bibinfo {author} {\bibfnamefont {B.~N.}\ \bibnamefont
  {Narozhny}}, \bibinfo {author} {\bibfnamefont {I.~V.}\ \bibnamefont
  {Gornyi}}, \bibinfo {author} {\bibfnamefont {M.}~\bibnamefont {Titov}},
  \bibinfo {author} {\bibfnamefont {M.}~\bibnamefont {Sch\"utt}}, \ and\
  \bibinfo {author} {\bibfnamefont {A.~D.}\ \bibnamefont {Mirlin}},\ }\bibfield
   {title} {\enquote {\bibinfo {title} {Hydrodynamics in graphene:
  Linear-response transport},}\ }\href {\doibase 10.1103/PhysRevB.91.035414}
  {\bibfield  {journal} {\bibinfo  {journal} {Phys. Rev. B}\ }\textbf {\bibinfo
  {volume} {91}},\ \bibinfo {pages} {035414} (\bibinfo {year}
  {2015})}\BibitemShut {NoStop}%
\bibitem [{\citenamefont {Narozhny}\ \emph {et~al.}(2017)\citenamefont
  {Narozhny}, \citenamefont {Gornyi}, \citenamefont {Mirlin},\ and\
  \citenamefont {Schmalian}}]{Schmalian:2017}%
  \BibitemOpen
  \bibfield  {author} {\bibinfo {author} {\bibfnamefont {Boris~N.}\
  \bibnamefont {Narozhny}}, \bibinfo {author} {\bibfnamefont {Igor~V.}\
  \bibnamefont {Gornyi}}, \bibinfo {author} {\bibfnamefont {Alexander~D.}\
  \bibnamefont {Mirlin}}, \ and\ \bibinfo {author} {\bibfnamefont {J{\"o}rg}\
  \bibnamefont {Schmalian}},\ }\bibfield  {title} {\enquote {\bibinfo {title}
  {Hydrodynamic approach to electronic transport in graphene},}\ }\href
  {\doibase https://doi.org/10.1002/andp.201700043} {\bibfield  {journal}
  {\bibinfo  {journal} {Annalen der Physik}\ }\textbf {\bibinfo {volume}
  {529}},\ \bibinfo {pages} {1700043} (\bibinfo {year} {2017})}\BibitemShut
  {NoStop}%
\bibitem [{\citenamefont {Levchenko}\ \emph {et~al.}(2024)\citenamefont
  {Levchenko}, \citenamefont {Li},\ and\ \citenamefont
  {Andreev}}]{Levchenko:2024}%
  \BibitemOpen
  \bibfield  {author} {\bibinfo {author} {\bibfnamefont {Alex}\ \bibnamefont
  {Levchenko}}, \bibinfo {author} {\bibfnamefont {Songci}\ \bibnamefont {Li}},
  \ and\ \bibinfo {author} {\bibfnamefont {A.~V.}\ \bibnamefont {Andreev}},\
  }\bibfield  {title} {\enquote {\bibinfo {title} {Giant magnetoresistance in
  weakly disordered non-{G}alilean invariant conductors},}\ }\href {\doibase
  10.1103/PhysRevB.109.075401} {\bibfield  {journal} {\bibinfo  {journal}
  {Phys. Rev. B}\ }\textbf {\bibinfo {volume} {109}},\ \bibinfo {pages}
  {075401} (\bibinfo {year} {2024})}\BibitemShut {NoStop}%
\bibitem [{\citenamefont {Xin}\ \emph {et~al.}(2023)\citenamefont {Xin},
  \citenamefont {Lourembam}, \citenamefont {Kumaravadivel}, \citenamefont
  {Kazantsev}, \citenamefont {Wu}, \citenamefont {Mullan}, \citenamefont
  {Barrier}, \citenamefont {Geim}, \citenamefont {Grigorieva}, \citenamefont
  {Mishchenko}, \citenamefont {Principi}, \citenamefont {Fal'ko}, \citenamefont
  {Ponomarenko}, \citenamefont {Geim},\ and\ \citenamefont
  {Berdyugin}}]{Ponomarenko:2023}%
  \BibitemOpen
  \bibfield  {author} {\bibinfo {author} {\bibfnamefont {Na}~\bibnamefont
  {Xin}}, \bibinfo {author} {\bibfnamefont {James}\ \bibnamefont {Lourembam}},
  \bibinfo {author} {\bibfnamefont {Piranavan}\ \bibnamefont {Kumaravadivel}},
  \bibinfo {author} {\bibfnamefont {A.~E.}\ \bibnamefont {Kazantsev}}, \bibinfo
  {author} {\bibfnamefont {Zefei}\ \bibnamefont {Wu}}, \bibinfo {author}
  {\bibfnamefont {Ciaran}\ \bibnamefont {Mullan}}, \bibinfo {author}
  {\bibfnamefont {Julien}\ \bibnamefont {Barrier}}, \bibinfo {author}
  {\bibfnamefont {Alexandra~A.}\ \bibnamefont {Geim}}, \bibinfo {author}
  {\bibfnamefont {I.~V.}\ \bibnamefont {Grigorieva}}, \bibinfo {author}
  {\bibfnamefont {A.}~\bibnamefont {Mishchenko}}, \bibinfo {author}
  {\bibfnamefont {A.}~\bibnamefont {Principi}}, \bibinfo {author}
  {\bibfnamefont {V.~I.}\ \bibnamefont {Fal'ko}}, \bibinfo {author}
  {\bibfnamefont {L.~A.}\ \bibnamefont {Ponomarenko}}, \bibinfo {author}
  {\bibfnamefont {A.~K.}\ \bibnamefont {Geim}}, \ and\ \bibinfo {author}
  {\bibfnamefont {Alexey~I.}\ \bibnamefont {Berdyugin}},\ }\bibfield  {title}
  {\enquote {\bibinfo {title} {Giant magnetoresistance of {D}irac plasma in
  high-mobility graphene},}\ }\href {\doibase 10.1038/s41586-023-05807-0}
  {\bibfield  {journal} {\bibinfo  {journal} {Nature}\ }\textbf {\bibinfo
  {volume} {616}},\ \bibinfo {pages} {270--274} (\bibinfo {year}
  {2023})}\BibitemShut {NoStop}%
\bibitem [{\citenamefont {Landau}\ and\ \citenamefont
  {Lifshitz}(1987)}]{LL-V6}%
  \BibitemOpen
  \bibfield  {author} {\bibinfo {author} {\bibfnamefont {L.~D.}\ \bibnamefont
  {Landau}}\ and\ \bibinfo {author} {\bibfnamefont {E.~M.}\ \bibnamefont
  {Lifshitz}},\ }\href@noop {} {\emph {\bibinfo {title} {Fluid Mechanics}}},\
  \bibinfo {edition} {2nd}\ ed.,\ \bibinfo {series} {Course of Theoretical
  Physics Series}, Vol.~\bibinfo {volume} {6}\ (\bibinfo  {publisher}
  {Butterworth-Heinemann, Oxford},\ \bibinfo {year} {1987})\BibitemShut
  {NoStop}%
\bibitem [{\citenamefont {Landau}\ and\ \citenamefont
  {Lifshitz}(1984)}]{LL-V8}%
  \BibitemOpen
  \bibfield  {author} {\bibinfo {author} {\bibfnamefont {L.~D.}\ \bibnamefont
  {Landau}}\ and\ \bibinfo {author} {\bibfnamefont {E.~M.}\ \bibnamefont
  {Lifshitz}},\ }\href@noop {} {\emph {\bibinfo {title} {Electrodynamics Of
  Continuous Media}}},\ \bibinfo {edition} {2nd}\ ed.,\ \bibinfo {series}
  {Course of Theoretical Physics Series}, Vol.~\bibinfo {volume} {8}\ (\bibinfo
   {publisher} {Butterworth-Heinemann},\ \bibinfo {year} {1984})\BibitemShut
  {NoStop}%
\bibitem [{\citenamefont {M\"uller}\ \emph {et~al.}(2009)\citenamefont
  {M\"uller}, \citenamefont {Schmalian},\ and\ \citenamefont
  {Fritz}}]{Muller:2009}%
  \BibitemOpen
  \bibfield  {author} {\bibinfo {author} {\bibfnamefont {Markus}\ \bibnamefont
  {M\"uller}}, \bibinfo {author} {\bibfnamefont {J\"org}\ \bibnamefont
  {Schmalian}}, \ and\ \bibinfo {author} {\bibfnamefont {Lars}\ \bibnamefont
  {Fritz}},\ }\bibfield  {title} {\enquote {\bibinfo {title} {Graphene: A
  nearly perfect fluid},}\ }\href {\doibase 10.1103/PhysRevLett.103.025301}
  {\bibfield  {journal} {\bibinfo  {journal} {Phys. Rev. Lett.}\ }\textbf
  {\bibinfo {volume} {103}},\ \bibinfo {pages} {025301} (\bibinfo {year}
  {2009})}\BibitemShut {NoStop}%
\bibitem [{\citenamefont {Avron}(1998)}]{Avron:1998}%
  \BibitemOpen
  \bibfield  {author} {\bibinfo {author} {\bibfnamefont {J.~E.}\ \bibnamefont
  {Avron}},\ }\bibfield  {title} {\enquote {\bibinfo {title} {Odd viscosity},}\
  }\href {\doibase 10.1023/a:1023084404080} {\bibfield  {journal} {\bibinfo
  {journal} {J. Stat. Phys.}\ }\textbf {\bibinfo {volume} {92}},\ \bibinfo
  {pages} {543--557} (\bibinfo {year} {1998})}\BibitemShut {NoStop}%
\bibitem [{\citenamefont {Torre}\ \emph {et~al.}(2015)\citenamefont {Torre},
  \citenamefont {Tomadin}, \citenamefont {Geim},\ and\ \citenamefont
  {Polini}}]{Torre:2015}%
  \BibitemOpen
  \bibfield  {author} {\bibinfo {author} {\bibfnamefont {Iacopo}\ \bibnamefont
  {Torre}}, \bibinfo {author} {\bibfnamefont {Andrea}\ \bibnamefont {Tomadin}},
  \bibinfo {author} {\bibfnamefont {Andre~K.}\ \bibnamefont {Geim}}, \ and\
  \bibinfo {author} {\bibfnamefont {Marco}\ \bibnamefont {Polini}},\ }\bibfield
   {title} {\enquote {\bibinfo {title} {Nonlocal transport and the hydrodynamic
  shear viscosity in graphene},}\ }\href {\doibase 10.1103/PhysRevB.92.165433}
  {\bibfield  {journal} {\bibinfo  {journal} {Phys. Rev. B}\ }\textbf {\bibinfo
  {volume} {92}},\ \bibinfo {pages} {165433} (\bibinfo {year}
  {2015})}\BibitemShut {NoStop}%
\bibitem [{\citenamefont {Levitov}\ and\ \citenamefont
  {Falkovich}(2016)}]{Levitov:2016}%
  \BibitemOpen
  \bibfield  {author} {\bibinfo {author} {\bibfnamefont {Leonid}\ \bibnamefont
  {Levitov}}\ and\ \bibinfo {author} {\bibfnamefont {Gregory}\ \bibnamefont
  {Falkovich}},\ }\bibfield  {title} {\enquote {\bibinfo {title} {Electron
  viscosity, current vortices and negative nonlocal resistance in graphene},}\
  }\href {\doibase 10.1038/nphys3667} {\bibfield  {journal} {\bibinfo
  {journal} {Nature Physics}\ }\textbf {\bibinfo {volume} {12}},\ \bibinfo
  {pages} {672--676} (\bibinfo {year} {2016})}\BibitemShut {NoStop}%
\bibitem [{\citenamefont {Alekseev}\ \emph {et~al.}(2018)\citenamefont
  {Alekseev}, \citenamefont {Dmitriev}, \citenamefont {Gornyi}, \citenamefont
  {Kachorovskii}, \citenamefont {Narozhny},\ and\ \citenamefont
  {Titov}}]{Alekseev:2018}%
  \BibitemOpen
  \bibfield  {author} {\bibinfo {author} {\bibfnamefont {P.~S.}\ \bibnamefont
  {Alekseev}}, \bibinfo {author} {\bibfnamefont {A.~P.}\ \bibnamefont
  {Dmitriev}}, \bibinfo {author} {\bibfnamefont {I.~V.}\ \bibnamefont
  {Gornyi}}, \bibinfo {author} {\bibfnamefont {V.~Yu.}\ \bibnamefont
  {Kachorovskii}}, \bibinfo {author} {\bibfnamefont {B.~N.}\ \bibnamefont
  {Narozhny}}, \ and\ \bibinfo {author} {\bibfnamefont {M.}~\bibnamefont
  {Titov}},\ }\bibfield  {title} {\enquote {\bibinfo {title} {Nonmonotonic
  magnetoresistance of a two-dimensional viscous electron-hole fluid in a
  confined geometry},}\ }\href {\doibase 10.1103/PhysRevB.97.085109} {\bibfield
   {journal} {\bibinfo  {journal} {Phys. Rev. B}\ }\textbf {\bibinfo {volume}
  {97}},\ \bibinfo {pages} {085109} (\bibinfo {year} {2018})}\BibitemShut
  {NoStop}%
\bibitem [{\citenamefont {Danz}\ \emph {et~al.}(2020)\citenamefont {Danz},
  \citenamefont {Titov},\ and\ \citenamefont {Narozhny}}]{Danz:2020}%
  \BibitemOpen
  \bibfield  {author} {\bibinfo {author} {\bibfnamefont {S.}~\bibnamefont
  {Danz}}, \bibinfo {author} {\bibfnamefont {M.}~\bibnamefont {Titov}}, \ and\
  \bibinfo {author} {\bibfnamefont {B.~N.}\ \bibnamefont {Narozhny}},\
  }\bibfield  {title} {\enquote {\bibinfo {title} {Giant nonlocality in nearly
  compensated two-dimensional semimetals},}\ }\href {\doibase
  10.1103/PhysRevB.102.081114} {\bibfield  {journal} {\bibinfo  {journal}
  {Phys. Rev. B}\ }\textbf {\bibinfo {volume} {102}},\ \bibinfo {pages}
  {081114} (\bibinfo {year} {2020})}\BibitemShut {NoStop}%
\bibitem [{\citenamefont {Mahan}\ \emph {et~al.}(1997)\citenamefont {Mahan},
  \citenamefont {Sales},\ and\ \citenamefont {Sharp}}]{Mahan:1997}%
  \BibitemOpen
  \bibfield  {author} {\bibinfo {author} {\bibfnamefont {Gerald}\ \bibnamefont
  {Mahan}}, \bibinfo {author} {\bibfnamefont {Brian}\ \bibnamefont {Sales}}, \
  and\ \bibinfo {author} {\bibfnamefont {Jeff}\ \bibnamefont {Sharp}},\
  }\bibfield  {title} {\enquote {\bibinfo {title} {Thermoelectric materials:
  New approaches to an old problem},}\ }\href {\doibase 10.1063/1.881752}
  {\bibfield  {journal} {\bibinfo  {journal} {Physics Today}\ }\textbf
  {\bibinfo {volume} {50}},\ \bibinfo {pages} {42--47} (\bibinfo {year}
  {1997})}\BibitemShut {NoStop}%
\bibitem [{\citenamefont {Messica}\ \emph {et~al.}(2023)\citenamefont
  {Messica}, \citenamefont {Ostrovsky},\ and\ \citenamefont
  {Gutman}}]{Messica:2023}%
  \BibitemOpen
  \bibfield  {author} {\bibinfo {author} {\bibfnamefont {Yonatan}\ \bibnamefont
  {Messica}}, \bibinfo {author} {\bibfnamefont {Pavel~M.}\ \bibnamefont
  {Ostrovsky}}, \ and\ \bibinfo {author} {\bibfnamefont {Dmitri~B.}\
  \bibnamefont {Gutman}},\ }\bibfield  {title} {\enquote {\bibinfo {title}
  {Heat transport in {W}eyl semimetals in the hydrodynamic regime},}\ }\href
  {\doibase 10.1103/PhysRevB.107.235102} {\bibfield  {journal} {\bibinfo
  {journal} {Phys. Rev. B}\ }\textbf {\bibinfo {volume} {107}},\ \bibinfo
  {pages} {235102} (\bibinfo {year} {2023})}\BibitemShut {NoStop}%
\end{thebibliography}%

\end{document}